\documentclass[english,intlimits,longbibliography,nofootinbib,10pt]{revtex4-2}
\usepackage{lmodern}

\usepackage[T1]{fontenc}
\usepackage[latin9]{inputenc}
\usepackage{color}
\usepackage{babel}
\usepackage{amsmath}
\usepackage{amssymb}
\usepackage{graphicx}
\usepackage[unicode=true,pdfusetitle,
 bookmarks=true,bookmarksnumbered=false,bookmarksopen=false,
 breaklinks=false,pdfborder={0 0 0},pdfborderstyle={},backref=false,colorlinks=true]
 {hyperref}

\makeatletter
\usepackage{graphicx}
\usepackage{amsmath}
\hypersetup{colorlinks,allcolors=blue,breaklinks=true}

\makeatother

\begin{document}
\title{Exact asymptotics of long-range quantum correlations in a nonequilibrium
steady state}
\author{Shachar Fraenkel}
\email{shacharf@mail.tau.ac.il}

\author{Moshe Goldstein}
\affiliation{Raymond and Beverly Sackler School of Physics and Astronomy, Tel Aviv
University, Tel Aviv 6997801, Israel}
\begin{abstract}
Out-of-equilibrium states of many-body systems tend to evade a description
by standard statistical mechanics, and their uniqueness is epitomized
by the possibility of certain long-range correlations that cannot
occur in equilibrium. In quantum many-body systems, coherent correlations
of this sort may lead to the emergence of remarkable entanglement
structures. In this work, we analytically study the asymptotic scaling
of quantum correlation measures -- the mutual information and the
fermionic negativity -- within the zero-temperature steady state
of voltage-biased free fermions on a one-dimensional lattice containing
a noninteracting impurity. Previously, we have shown that two subsystems
on opposite sides of the impurity exhibit volume-law entanglement,
which is independent of the absolute distances of the subsystems from
the impurity. Here we go beyond that result and derive the exact form
of the subleading logarithmic corrections to the extensive terms of
correlation measures, in excellent agreement with numerical calculations.
In particular, the logarithmic term of the mutual information asymptotics
can be encapsulated in a concise formula, depending only on simple
four-point ratios of subsystem length-scales and on the impurity scattering
probabilities at the Fermi energies. This echoes the case of equilibrium
states, where such logarithmic terms may convey universal information
about the physical system. To compute these exact results, we devise
a hybrid method that relies on Toeplitz determinant asymptotics for
correlation matrices in both real space and momentum space, successfully
circumventing the inhomogeneity of the system. This method can potentially
find wider use for analytical calculations of entanglement measures
in similar scenarios.
\end{abstract}
\maketitle


\tableofcontents

\section{Introduction}

When attempting to paint a universal picture of condensed matter systems,
the study of correlations is by far one of the most potent and versatile
brushes at hand. For a rich palette of both classical and quantum
many-body models, the spatial range of correlations is the defining
feature of their distinct phases of matter. Furthermore, quantitative
correlation measures can be used to unify the description of phase
transitions, by identifying common types of functional dependence
of these measures on model parameters or on thermodynamic variables
\citep{cardy_1996,sachdev_2011,coleman_2015}.

In particular, this framework particularly permits to incorporate
out-of-equilibrium many-body systems, for which the traditional statistical
mechanics description fails, into the same universal picture. The
boundaries of this picture must, however, be expanded for this purpose:
the stationary states of open systems may deviate drastically in their
characteristics from equilibrium states, as can be usefully captured
by the long-range correlations that they tend to exhibit \citep{Derrida_2007,PhysRevLett.123.080601,PhysRevX.13.011045}.
For quantum systems, an obvious emphasis should be put on analyzing
measures of many-body entanglement \citep{RevModPhys.80.517,LAFLORENCIE20161},
as they embody specifically the non-classical aspect of this departure
from the familiar landscape of equilibrium \citep{PhysRevA.89.032321,Eisler_2014,PhysRevB.96.054302,PhysRevLett.123.110601,PhysRevX.9.021007,PhysRevB.101.180301,PhysRevB.103.035108,PhysRevB.103.L020302,carollo2021emergent,10.21468/SciPostPhys.11.4.085,10.21468/SciPostPhys.12.1.011,alba2022logarithmic,murciano2023symmetryresolved,bernard2023exact}.
The promise of the entanglement perspective is underlined by the significant
achievements that it has already produced in the classification of
quantum phases, of topological phases, and of dynamical traits of
closed quantum systems after a nonequilibrium quench \citep{PhysRevA.66.032110,PhysRevLett.101.010504,doi:10.1080/00018732.2016.1198134,doi:10.1073/pnas.1703516114,RevModPhys.91.021001,Serbyn2021}.

When studying entanglement in extended many-body systems, the standard
modus operandi includes calculating measures of entanglement between
generic subsystems, and observing the asymptotic scaling of these
measures with the sizes of the subsystems and the distance between
them. Universality may be present both in the leading-order asymptotics,
as well as in subleading terms \citep{Calabrese_2004,PhysRevLett.96.010404,PhysRevLett.96.110404,RevModPhys.82.277}.
This observation, by now firmly entrenched thanks to numerous studies,
has been driving the need for the development of analytical techniques
suited for performing such asymptotic calculations. Moreover, while
the initial focus of this vast body of work has been naturally put
on homogeneous systems, a recent surge of interest is directed toward
the nontrivial signatures that inhomogeneities (e.g., boundaries and
defects) may imprint on correlation and entanglement structures, either
in equilibrium \citep{PhysRevB.88.085413,PhysRevLett.128.090603,Mintchev2022,horvath2023chargeresolved,CapizziFCS_2023}
or out of equilibrium \citep{Eisler_2012,10.21468/SciPostPhys.6.1.004,Gruber_2020,Capizzi_2023,Gouraud_2023,10.21468/SciPostPhys.14.4.070,rylands2023transport}.
This avenue of research raises particular technical challenges when
one attempts to adapt well-established analytical methods that had
originally relied on the translation invariance of the system in question.

Building on this general theme, in this work we consider the correlation
structure of the nonequilibrium steady state of an open one-dimensional
free fermion chain, where the homogeneity of the chain is broken by
a noninteracting impurity. Specifically, we treat the case where the
system is biased by two zero-temperature reservoirs with different
chemical potentials, and discuss an arbitrary current-conserving impurity,
such that our analysis requires only the knowledge of its associated
scattering matrix. We examine the correlations between two subsystems
of the chain located on opposite sides of the impurity, as quantified
by the mutual information (MI) and the fermionic negativity (both
measures will be defined precisely in Sec.~\ref{sec:Preliminaries}).

The analysis presented here extends our recent work \citep{fraenkel2022extensive},
where we considered the same scenario and calculated the leading-order
asymptotics of the MI and the negativity. There, we observed that
both the MI and the negativity scale linearly with the overlap between
one subsystem and the mirror image of the other subsystem, obtained
from the reflection of the latter subsystem about the position of
the impurity (see Sec.~\ref{sec:Preliminaries} for full details
on the results derived in Ref.~\citep{fraenkel2022extensive}). This
result thus describes a volume-law scaling of correlations (and, in
particular, of entanglement), which does not decay with the distance
between the two subsystems (as long as the size of the mirror-image
overlap is kept constant), in stark contrast to the known behavior
of equilibrium states.

Here we go beyond that leading-order asymptotics and derive the first
subleading corrections to the MI and negativity, which turn out to
be logarithmic in the different length scales of the problem (i.e.,
the lengths of the subsystems and their distances from the impurity).
These corrections arise from discontinuous jumps in the local distribution
of single-particle momentum states, jumps which in turn stem from
the fact that the edge reservoirs are at zero temperature. The logarithmic
scaling of entanglement measures is indeed a hallmark of 1D quantum
critical many-body states \citep{PhysRevLett.90.227902,Paul2022Hidden},
and the associated asymptotic scaling coefficients are known to encode
fundamental properties of models giving rise to such states, e.g.,
the topology of the Fermi sea in a Fermi liquid \citep{PhysRevX.12.031022},
or the central charge in a conformal field theory \citep{Calabrese_2004,Calabrese_2009}.

In order to perform the calculation, we employ and extend analytical
techniques related to the asymptotic scaling of Toeplitz determinants,
which is given by the Fisher-Hartwig formula \citep{Jin2004,10.2307/23030524}.
While these techniques are most commonly used when considering correlations
within translation-invariant models, we show how they can in fact
be utilized to capture the effects of broken homogeneity on steady-state
correlations. Our methodology uses separate insights from particular
cases that are carefully chosen, where the two-point correlation matrix
has a Toeplitz (or block-Toeplitz) structure either in real space
or in momentum space, and sews them together to obtain our general
results. The results we report here may therefore prove to be valuable
on the technical level as well as on the fundamental level.

The remainder of the paper is organized as follows. In Sec.~\ref{sec:Preliminaries}
we set the stage by recalling the basic definitions for the correlation
measures of interest, and by introducing the specific model that we
examine, as well as several useful notations. We also recap relevant
previous results from Refs.~\citep{10.21468/SciPostPhys.11.4.085}
and \citep{fraenkel2022extensive}. In Sec.~\ref{sec:Asymptotics-beyond-volume-law}
we state our main analytical results for the asymptotics of the MI
and the negativity beyond the leading-order volume law, and present
comparisons of these analytical results to results obtained from numerical
computations (subtler details of these comparisons are also discussed
in Appendix \ref{sec:Numerics-comparison-appendix}). Sec.~\ref{sec:Derivation-of-results}
details the method used to derive our analytical results, while we
defer the discussion of several more technical aspects to three appendices
(\ref{sec:Stationary-phase-approximation}--\ref{sec:Negativity-derivation-appendix})
at the end of paper. Sec.~\ref{sec:Discussion} includes a concluding
discussion of our analysis.

\section{Preliminaries\label{sec:Preliminaries}}

In this preparatory section, we present the definitions of the quantum
correlation measures that will be used throughout our analysis (Subsec.~\ref{subsec:Correlation-measures-Definitions});
describe the model and its nonequilibrium steady state on which this
paper focuses (Subsec.~\ref{subsec:Model}); review pertinent results
that were previously derived in Refs.~\citep{10.21468/SciPostPhys.11.4.085}
and \citep{fraenkel2022extensive} (Subsec.~\ref{subsec:Recap});
and set up notations for several functions that arise in our analytical
results (Subsec.~\ref{subsec:Preliminary-notations}).

\subsection{Correlation measures: Definitions\label{subsec:Correlation-measures-Definitions}}

We begin by reviewing the definitions of several well-known quantitative
measures of quantum correlations. The $n$th R\'enyi entropy of a
subsystem $X$ is defined as
\begin{equation}
S_{X}^{\left(n\right)}=\frac{1}{1-n}\ln{\rm Tr}\!\left[\left(\rho_{{\scriptscriptstyle X}}\right)^{n}\right],
\end{equation}
where $\rho_{{\scriptscriptstyle X}}$ is the reduced density matrix
of $X$, obtained by tracing out the degrees of freedom of the complementary
subsystem from the state of the full system. The R\'enyi entropies
can be analytically continued to obtain the von Neumann entropy, given
by
\begin{equation}
{\cal S}_{X}=\lim_{n\to1}S_{X}^{\left(n\right)}=-{\rm Tr}\!\left[\rho_{{\scriptscriptstyle X}}\ln\rho_{{\scriptscriptstyle X}}\right].
\end{equation}
When considering a system in a pure state, the von Neumman entropy
${\cal S}_{X}$ is the optimal measure quantifying the entanglement
of subsystem $X$ with its complement \citep{RevModPhys.81.865}.
The calculation of R\'enyi entropies provides a way to obtain the
von Neumann entropy, but beyond that they hold importance as entanglement
monotones, which are also easier to extract through experimental measurements
than the von Neumann entropy~\citep{PhysRevLett.109.020504,PhysRevLett.109.020505,Islam2015,PhysRevLett.120.050406,PhysRevA.99.062309}.
Furthermore, full knowledge of the R\'enyi entropies enables the
reconstruction of the entanglement spectrum \citep{PhysRevA.78.032329}.

To quantify the correlations between two arbitrary subsystems $X_{1}$
and $X_{2}$, one may use their mutual information (MI), defined as
the following combination of von Neumann entropies:
\begin{equation}
{\cal I}_{X_{1}:X_{2}}={\cal S}_{X_{1}}+{\cal S}_{X_{2}}-{\cal S}_{X_{1}\cup X_{2}}.\label{eq:vN-MI-definition}
\end{equation}
The MI encodes the total amount of correlation between the two subsystems,
counting both classical and quantum correlations indistinguishably~\citep{PhysRevA.72.032317}.
Again one may define a corresponding R\'enyi quantity, combining
simple moments:
\begin{equation}
{\cal I}_{X_{1}:X_{2}}^{\left(n\right)}=S_{X_{1}}^{\left(n\right)}+S_{X_{2}}^{\left(n\right)}-S_{X_{1}\cup X_{2}}^{\left(n\right)}.\label{eq:Renyi-MI-definition}
\end{equation}
Under analytic continuation, the R\'enyi MI yields the von Neumann
MI through ${\cal I}_{X_{1}:X_{2}}=\lim_{n\to1}{\cal I}_{X_{1}:X_{2}}^{\left(n\right)}$.

For the purpose of capturing the quantum correlations alone, a convenient
measure is given by the fermionic negativity~\citep{PhysRevB.95.165101}
\begin{equation}
{\cal E}=\ln{\rm Tr}\sqrt{\left(\widetilde{\rho}_{{\scriptscriptstyle X_{1}\cup X_{2}}}\right)^{\dagger}\widetilde{\rho}_{{\scriptscriptstyle X_{1}\cup X_{2}}}},
\end{equation}
where here $\widetilde{\rho}_{{\scriptscriptstyle X_{1}\cup X_{2}}}$
is the reduced density matrix of $X_{1}\cup X_{2}$ following a partial
time-reversal transformation (i.e., a time-reversal transformation
applied to either $X_{1}$ or $X_{2}$). The fermionic negativity
is an entanglement monotone \citep{PhysRevA.99.022310} which is a
variant of the more widely used logarithmic negativity \citep{PhysRevA.65.032314,PhysRevLett.95.090503},
and it is more suitable than the latter for treating fermionic Gaussian
states (like the state discussed in this work), given that the partial
time-reversal preserves fermionic Gaussianity \citep{PhysRevB.95.165101,PhysRevB.97.165123}.

Just like the von Neumann entropy, the fermionic negativity may be
expressed as the analytic continuation of simpler moments; namely,
if we define for any even integer $n$ the R\'enyi negativity
\begin{equation}
{\cal E}_{n}=\ln{\rm Tr}\!\left[\left(\left(\widetilde{\rho}_{{\scriptscriptstyle X_{1}\cup X_{2}}}\right)^{\dagger}\widetilde{\rho}_{{\scriptscriptstyle X_{1}\cup X_{2}}}\right)^{n/2}\right],
\end{equation}
then ${\cal E}$ is the outcome of its analytic continuation to $n=1$.
Note that, given a separable state of $X_{1}\cup X_{2}$, the matrices
$\rho_{{\scriptscriptstyle X_{1}\cup X_{2}}}$ and $\widetilde{\rho}_{{\scriptscriptstyle X_{1}\cup X_{2}}}$
have identical spectra \citep{PhysRevA.99.022310}, meaning that ${\cal E}_{n}=\left(1-n\right)S_{X_{1}\cup X_{2}}^{\left(n\right)}$.
This entails that, whenever $X_{1}\cup X_{2}$ is in a mixed state,
the R\'enyi negativities are non-zero even in the absence of any
correlations between $X_{1}$ and $X_{2}$. Thus, R\'enyi negativities
cannot be regarded as direct measures of correlations, but they can
nevertheless be instrumental in extracting the fermionic negativity
or the negativity spectrum \citep{PhysRevB.94.195121,10.21468/SciPostPhys.7.3.037}.

\subsection{\label{subsec:Model}Model}

The model we consider in this work is that of free fermions hopping
on a one-dimensional infinite lattice, described by the tight-binding
Hamiltonian
\begin{equation}
{\cal H}=-\eta\!\sum_{m=m_{0}}^{\infty}\!\!\left[c_{m}^{\dagger}c_{m+1}+c_{-m}^{\dagger}c_{-m-1}+{\rm h.c.}\right]+{\cal H}_{{\rm scat}},\label{eq:Model-Hamiltonian}
\end{equation}
where $\eta>0$ is the hopping amplitude and $c_{m}$ is the fermionic
annihilation operator for site $m$. The homogeneity of the Hamiltonian
is broken by the noninteracting (quadratic) term ${\cal H}_{{\rm scat}}$,
which represents the current-conserving impurity that spans only the
sites $\left|m\right|\le m_{0}$ near the middle of the chain, such
that $2m_{0}+1$ can be regarded as the size of the impurity region.
In general, ${\cal H}_{{\rm scat}}$ is a linear combination of terms
of the form $c_{m}^{\dagger}c_{m'}$ with $\left|m\right|,\left|m'\right|\le m_{0}$,
or possibly with $c_{m}^{\dagger}$ or $c_{m'}$ being replaced with
the corresponding operator of a side-attached site.

In all, the impurity has a minor effect on the single-particle energy
spectrum compared to that of the homogeneous tight-binding model,
but a dramatic effect on the wavefunctions of the energy eigenstates.
Instead of momentum eigenstates with plane-wave wavefunctions, for
$0<k<\pi$ we can define eigenstates $|k\rangle$, with their wavefunctions
given by
\begin{equation}
\left\langle m|k\right\rangle =\begin{cases}
e^{ikm}+r_{{\scriptscriptstyle L}}\!\left(\left|k\right|\right)e^{-ikm} & m<-m_{0},\\
t_{{\scriptscriptstyle L}}\!\left(\left|k\right|\right)e^{ikm} & m>m_{0}.
\end{cases}\label{eq:Left-scattering-states}
\end{equation}
We refer to these eigenstates as ``left scattering states''. For
$-\pi<k<0$ we define ``right scattering states'', the corresponding
eigenstate wavefunctions of which are
\begin{equation}
\left\langle m|k\right\rangle =\begin{cases}
t_{{\scriptscriptstyle R}}\!\left(\left|k\right|\right)e^{ikm} & m<-m_{0},\\
e^{ikm}+r_{{\scriptscriptstyle R}}\!\left(\left|k\right|\right)e^{-ikm} & m>m_{0}.
\end{cases}\label{eq:Right-scattering-states}
\end{equation}
The above wavefunctions feature reflection and transmission amplitudes,
which are determined by the particular properties of the impurity.
The labels $L$ and $R$ refer to the particle incoming from the left
or from the right (respectively) before being scattered off the impurity.
In Eqs.~(\ref{eq:Left-scattering-states})--(\ref{eq:Right-scattering-states})
we disregarded the form of the eigenstate wavefunctions inside the
impurity region, both because their form is not universal in this
region, and because this has no bearing on correlations between subsystems
outside this region (which are the subject of this work). The scattering
amplitudes can be organized in a so-called scattering matrix defined
for any $0<k<\pi$:
\begin{equation}
S\!\left(k\right)=\left(\begin{array}{cc}
r_{{\scriptscriptstyle L}}\!\left(k\right) & t_{{\scriptscriptstyle R}}\!\left(k\right)\\
t_{{\scriptscriptstyle L}}\!\left(k\right) & r_{{\scriptscriptstyle R}}\!\left(k\right)
\end{array}\right).\label{eq:Sacttering-matrix}
\end{equation}
The scattering matrix $S\!\left(k\right)$ is unitary \citep{merzbacher1998quantum}.
In particular, if we let ${\cal T}=\left|t_{{\scriptscriptstyle L}}\right|^{2}=\left|t_{{\scriptscriptstyle R}}\right|^{2}$
denote the transmission probability and ${\cal R}=\left|r_{{\scriptscriptstyle L}}\right|^{2}=\left|r_{{\scriptscriptstyle R}}\right|^{2}$
denote the reflection probability, we always have ${\cal T}\!\left(k\right)+{\cal R}\!\left(k\right)=1$.

The energy associated with each of the eigenstates in Eqs.~(\ref{eq:Left-scattering-states})--(\ref{eq:Right-scattering-states})
is determined by $k$, namely $\epsilon\!\left(k\right)=-2\eta\cos k$.
The single-particle energy eigenbasis generically contains bound states
in addition to these extended eigenstates; however, these bound states
are exponentially localized at the impurity \citep{merzbacher1998quantum,doi:10.1063/1.525968},
and thus their negligible effect on the correlations considered in
this work can be ignored.

\begin{figure}
\begin{centering}
\includegraphics[viewport=20bp 150bp 1260bp 450bp,clip,width=1\columnwidth]{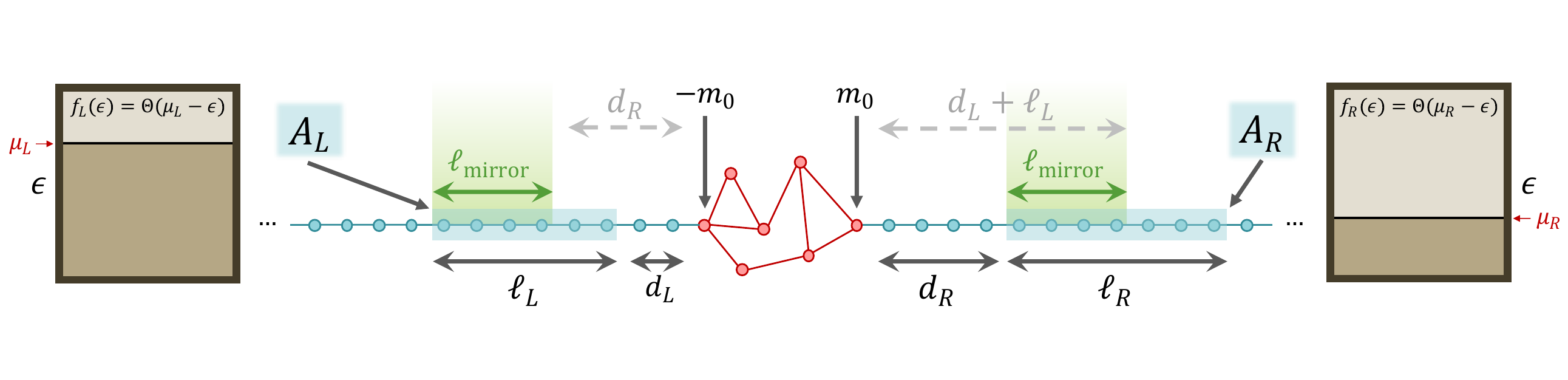}
\par\end{centering}
\caption{\label{fig:Schematic-illustration}Schematic illustration of the model,
comprised of a one-dimensional lattice that contains a generic current-conserving
noninteracting impurity, and is connected at its ends to two zero-temperature
reservoirs of noninteracting fermions. Sites that constitute the impurity
region are marked in red, and all other sites are marked in blue.
See Subsec.~\ref{subsec:Model} of the text for further details on
the notations, and Eq.~(\ref{eq:mirror-image-overlap-definition})
for the definition of $\ell_{{\rm mirror}}$.}
\end{figure}

The many-body nonequilibrium steady state that we inspect originates
in an occupation bias between left and right scattering states: left
scattering states are occupied up to a chemical potential $\mu_{{\scriptscriptstyle L}}$,
and right scattering states are occupied up to a chemical potential
$\mu_{{\scriptscriptstyle R}}$, with $\mu_{{\scriptscriptstyle L}}\neq\mu_{{\scriptscriptstyle R}}$.
We can define corresponding Fermi momenta through $k_{{\scriptscriptstyle F,i}}=\cos^{-1}\!\left(-\mu_{i}/2\eta\right)$
(for $i=L,R$), such that the many-body state is in essence a tilted
Fermi sea. This steady state can be thought of as the result of connecting
the chain to two edge reservoirs containing free fermions at zero
temperature and with the chemical potentials $\mu_{{\scriptscriptstyle L}}$
and $\mu_{{\scriptscriptstyle R}}$ (such that, in each reservoir,
single-particle energy states are occupied with probability $f_{i}\!\left(\epsilon\right)=\Theta\!\left(\mu_{i}-\epsilon\right)$
as a function of the energy $\epsilon$, with $\Theta$ being the
Heaviside step function), and subsequently going to the thermodynamic
limit of an infinite chain; such a scenario is depicted in Fig.~\ref{fig:Schematic-illustration}.
Alternatively, the steady state can be seen as the long-time limit
following a quench where two zero-temperature Fermi seas with the
chemical potentials $\mu_{{\scriptscriptstyle L}}$ and $\mu_{{\scriptscriptstyle R}}$
are joined together at the impurity.

The most remarkable correlation properties of this nonequilibrium
steady state are revealed when considering the correlations between
two subsystems on opposite sides of the impurity. We specifically
consider an interval $A_{{\scriptscriptstyle L}}$ to the left of
the impurity, containing the sites $m$ where $-d_{{\scriptscriptstyle L}}-\ell_{{\scriptscriptstyle L}}\le m+m_{0}\le-d_{{\scriptscriptstyle L}}-1$,
and another interval $A_{{\scriptscriptstyle R}}$ to the right of
the impurity, containing the sites $m$ where $d_{{\scriptscriptstyle R}}+1\le m-m_{0}\le d_{{\scriptscriptstyle R}}+\ell_{{\scriptscriptstyle R}}$.
For the subsystem $A_{i}$, the length scales $\ell_{i}$ and $d_{i}$
thus represent the length of the subsystem and its distance from the
impurity region, respectively. An example for such a configuration
is illustrated in Fig.~\ref{fig:Schematic-illustration}. For our
asymptotic calculation, we assume that these length scales are all
much larger than the size of the impurity, i.e., $\ell_{i},d_{i}\gg2m_{0}+1$.
For future convenience, we let $A=A_{{\scriptscriptstyle L}}\cup A_{{\scriptscriptstyle R}}$
denote the union of the two subsystems of interest.

\subsection{Recap of previous results\label{subsec:Recap}}

Here we recount the results reported in Refs.~\citep{10.21468/SciPostPhys.11.4.085,fraenkel2022extensive}
for entropies and correlation measures associated with subsystems
$A_{{\scriptscriptstyle L}}$ and $A_{{\scriptscriptstyle R}}$. As
already mentioned, the results of Ref.~\citep{fraenkel2022extensive}
included only the leading-order (volume-law) terms in the asymptotics
of these quantities. In Appendix \ref{sec:Stationary-phase-approximation}
we repeat a crucial technical step that was performed in Ref.~\citep{fraenkel2022extensive}
to obtain these results, since we rely on it again in our current
derivation of the subleading corrections.

Most notably, in Ref.~\citep{fraenkel2022extensive} we observed
that, to a leading order, all the aforementioned quantities of interest
(the MI, the negativity and their R\'enyi counterparts) depend on
the distances $d_{{\scriptscriptstyle L}}$ and $d_{{\scriptscriptstyle R}}$
from the impurity only through their difference $d_{{\scriptscriptstyle L}}-d_{{\scriptscriptstyle R}}$.
In particular, they remain the same when fixing this difference and
taking the limit $d_{i}/\ell_{i}\to\infty$, revealing strong long-range
correlations and entanglement in the steady state. Due to this behavior,
the results of Ref.~\citep{fraenkel2022extensive} are most conveniently
presented if one first defines
\begin{equation}
\ell_{{\rm mirror}}=\max\!\left\{ \min\!\left\{ d_{{\scriptscriptstyle L}}+\ell_{{\scriptscriptstyle L}},d_{{\scriptscriptstyle R}}+\ell_{{\scriptscriptstyle R}}\right\} -\max\!\left\{ d_{{\scriptscriptstyle L}},d_{{\scriptscriptstyle R}}\right\} ,0\right\} ,\label{eq:mirror-image-overlap-definition}
\end{equation}
which is the mirror-image overlap between subsystems $A_{{\scriptscriptstyle L}}$
and $A_{{\scriptscriptstyle R}}$: $\ell_{{\rm mirror}}$ counts the
number of pairs $\left(-m,m\right)\in A_{{\scriptscriptstyle L}}\times A_{{\scriptscriptstyle R}}$;
that is, the number of sites shared between the subsystems upon reflecting
one of them about $m=0$ (see Fig.~\ref{fig:Schematic-illustration}).
$\ell_{{\rm mirror}}$ is conspicuously dependent on $d_{{\scriptscriptstyle L}}$
and $d_{{\scriptscriptstyle R}}$ only through $d_{{\scriptscriptstyle L}}-d_{{\scriptscriptstyle R}}$.
In what follows, we will also use the notations $\Delta\ell_{i}=\ell_{i}-\ell_{{\rm mirror}}$,
$k_{+}=\max\!\left\{ k_{{\scriptscriptstyle F,L}},k_{{\scriptscriptstyle F,R}}\right\} $
and $k_{-}=\min\!\left\{ k_{{\scriptscriptstyle F,L}},k_{{\scriptscriptstyle F,R}}\right\} $.

To be explicit, in Ref.~\citep{fraenkel2022extensive} we have found
that, to a leading order, the R\'enyi entropies of subsystems $A_{{\scriptscriptstyle L}}$,
$A_{{\scriptscriptstyle R}}$ and $A$ are given by

\begin{align}
S_{A_{i}}^{\left(n\right)} & \sim\frac{\ell_{i}}{1-n}\int_{k_{-}}^{k_{+}}\frac{dk}{2\pi}\ln\!\left[\left({\cal T}\!\left(k\right)\right)^{n}+\left({\cal R}\!\left(k\right)\right)^{n}\right],\nonumber \\
S_{A}^{\left(n\right)} & \sim\frac{\Delta\ell_{{\scriptscriptstyle L}}+\Delta\ell_{{\scriptscriptstyle R}}}{1-n}\int_{k_{-}}^{k_{+}}\frac{dk}{2\pi}\ln\!\left[\left({\cal T}\!\left(k\right)\right)^{n}+\left({\cal R}\!\left(k\right)\right)^{n}\right].\label{eq:Renyi-entropies-asymptotics}
\end{align}
This directly leads (recall Eq.~(\ref{eq:Renyi-MI-definition}))
to the asymptotics of the R\'enyi MI, namely

\begin{equation}
{\cal I}_{A_{L}:A_{R}}^{\left(n\right)}\sim\frac{\ell_{{\rm mirror}}}{1-n}\!\int_{k_{-}}^{k_{+}}\frac{dk}{\pi}\ln\!\left[\left({\cal T}\!\left(k\right)\right)^{n}+\left({\cal R}\!\left(k\right)\right)^{n}\right],\label{eq:Renyi_MI_volume_law}
\end{equation}
and the MI is then simply obtained by taking the $n\to1$ limit:

\begin{equation}
{\cal I}_{A_{L}:A_{R}}\sim\ell_{{\rm mirror}}\!\int_{k_{-}}^{k_{+}}\frac{dk}{\pi}\left[-{\cal T}\!\left(k\right)\ln{\cal T}\!\left(k\right)-\!{\cal R}\!\left(k\right)\ln{\cal R}\!\left(k\right)\right].\label{eq:MI_volume_law}
\end{equation}
Furthermore, the R\'enyi negativities were found to satisfy

\begin{align}
{\cal E}_{n} & \sim\ell_{{\rm mirror}}\int_{k_{-}}^{k_{+}}\frac{dk}{\pi}\ln\!\left[\left({\cal T}\!\left(k\right)\right)^{n/2}+\left({\cal R}\!\left(k\right)\right)^{n/2}\right]+\left(\Delta\ell_{{\scriptscriptstyle L}}+\Delta\ell_{{\scriptscriptstyle R}}\right)\int_{k_{-}}^{k_{+}}\frac{dk}{2\pi}\ln\!\left[\left({\cal T}\!\left(k\right)\right)^{n}+\left({\cal R}\!\left(k\right)\right)^{n}\right],\label{eq:Renyi-negativities-asymptotics}
\end{align}
which in the limit $n\to1$ gives the behavior of the fermionic negativity:

\begin{equation}
{\cal E}\sim\ell_{{\rm mirror}}\!\int_{k_{-}}^{k_{+}}\frac{dk}{\pi}\ln\!\left[\left({\cal T}\!\left(k\right)\right)^{1/2}+\left({\cal R}\!\left(k\right)\right)^{1/2}\right].\label{eq:Negativity_volume_law}
\end{equation}

A salient attribute of Eqs.~(\ref{eq:Renyi-entropies-asymptotics})--(\ref{eq:Negativity_volume_law})
is that the volume-law terms appearing in all of them identically
vanish either in the absence of the bias (i.e., if $k_{{\scriptscriptstyle F,L}}=k_{{\scriptscriptstyle F,R}}$),
or if ${\cal T}\!\left(k\right)\in\left\{ 0,1\right\} $ for all $k\in\left[k_{-},k_{+}\right]$;
we refer to the latter scenario as that of a ``trivial'' impurity,
which either perfectly transmits or perfectly reflects incoming particles.
If instead both the bias and the scattering are nontrivial, then the
volume-law terms in Eqs.~(\ref{eq:Renyi-entropies-asymptotics})--(\ref{eq:Negativity_volume_law})
are nonzero. In other words, the voltage bias and the nontrivial scattering
are necessary and sufficient conditions for observing the extensive
entropies in Eq.~(\ref{eq:Renyi-entropies-asymptotics}) and the
extensive long-range correlations captured by Eqs.~(\ref{eq:Renyi_MI_volume_law})--(\ref{eq:Negativity_volume_law}).

Another important result on which we will rely in our analysis pertains
to the subleading logarithmic terms of the R\'enyi entropies for
a single interval $A_{i}$ in the long-range regime $d_{i}/\ell_{i}\gg1$.
In Ref.~\citep{10.21468/SciPostPhys.11.4.085} we have shown that,
for $d_{i}/\ell_{i}\gg1$ and up to ${\cal O}\!\left(1\right)$ corrections,
the single-interval entropies obey
\begin{align}
S_{A_{L}}^{\left(n\right)} & \sim\frac{\ell_{{\scriptscriptstyle L}}}{1-n}\int_{k_{-}}^{k_{+}}\frac{dk}{2\pi}\ln\!\left[\left({\cal T}\!\left(k\right)\right)^{n}+\left({\cal R}\!\left(k\right)\right)^{n}\right]+\frac{1+n}{12n}\ln\ell_{{\scriptscriptstyle L}}+\frac{\ln\ell_{{\scriptscriptstyle L}}}{1-n}\left[Q_{n}\!\left({\cal T}\!\left(k_{{\scriptscriptstyle F,L}}\right)\right)+Q_{n}\!\left({\cal R}\!\left(k_{{\scriptscriptstyle F,R}}\right)\right)\right],\nonumber \\
S_{A_{R}}^{\left(n\right)} & \sim\frac{\ell_{{\scriptscriptstyle R}}}{1-n}\int_{k_{-}}^{k_{+}}\frac{dk}{2\pi}\ln\!\left[\left({\cal T}\!\left(k\right)\right)^{n}+\left({\cal R}\!\left(k\right)\right)^{n}\right]+\frac{1+n}{12n}\ln\ell_{{\scriptscriptstyle R}}+\frac{\ln\ell_{{\scriptscriptstyle R}}}{1-n}\left[Q_{n}\!\left({\cal T}\!\left(k_{{\scriptscriptstyle F,R}}\right)\right)+Q_{n}\!\left({\cal R}\!\left(k_{{\scriptscriptstyle F,L}}\right)\right)\right],\label{eq:Renyi-entropies-single-intervals}
\end{align}
where the function $Q_{n}$ is defined below, see Eq.~(\ref{eq:Log-scaling-kernel-overlap}).

A word of caution is in order regarding Eq.~(\ref{eq:Renyi-entropies-single-intervals}):
its derivation assumes a finite bias ($k_{{\scriptscriptstyle F,L}}\neq k_{{\scriptscriptstyle F,R}}$),
and the result for the case of no bias ($k_{{\scriptscriptstyle F,L}}=k_{{\scriptscriptstyle F,R}}$)
cannot be recovered by naively taking the corresponding limit of Eq.~(\ref{eq:Renyi-entropies-single-intervals}).
This is because the logarithmic term arises from discontinuities in
the local distribution of momentum states, and in this no-bias limit
two discontinuities merge into a single one (this issue comes up again
in the derivation of the logarithmic corrections to $S_{A}^{\left(n\right)}$,
and so we discuss it in more technical detail in Subsec.~\ref{subsec:MI-derivation-Symmetric-configuration}).
The proper limit of the entropies in the no-bias case is instead $S_{A_{i}}^{\left(n\right)}\sim\frac{1+n}{6n}\ln\ell_{i}$,
equal to the equilibrium result for free fermions on a homogeneous
tight-binding lattice \citep{Calabrese_2004}.

\subsection{Useful notations\label{subsec:Preliminary-notations}}

Here we introduce several useful notations that will make the presentation
of our main results more concise. For any $0\le p\le1$ and any $n$,
we define the function
\begin{align}
Q_{n}\!\left(p\right) & =\frac{n}{2\pi^{2}}\int_{p}^{1}dx\frac{x^{n-1}-\left(1-x\right)^{n-1}}{x^{n}+\left(1-x\right)^{n}}\ln\!\left(\frac{1-x}{x-p}\right)\nonumber \\
 & =-\frac{n}{12}+\int_{0}^{1}\frac{dx}{2\pi^{2}x}\left\{ \ln\!\left[\left(1+px\right)^{n}+\left(\left(1-p\right)x\right)^{n}\right]+\ln\!\left[\frac{\left(x+p\right)^{n}+\left(1-p\right)^{n}}{p^{n}+\left(1-p\right)^{n}}\right]\right\} .\label{eq:Log-scaling-kernel-overlap}
\end{align}
This function in particular satisfies $Q_{n}\!\left(1\right)=0$ and
$Q_{n}\!\left(0\right)=\frac{1}{12}\left(\frac{1}{n}-n\right)$. Note
that the latter is (up to a minus sign) the well-known scaling dimension
of a branch-point twist field for an equilibrium free-fermion field
theory, assuming that this theory is defined on the $n$-sheet Riemann
surface that is conventionally used (as part of the so-called ``replica
trick'') in conformal field theory calculations of entropy and entanglement
measures \citep{Calabrese_2009}.

\begin{figure}
\begin{centering}
\includegraphics[width=0.45\columnwidth]{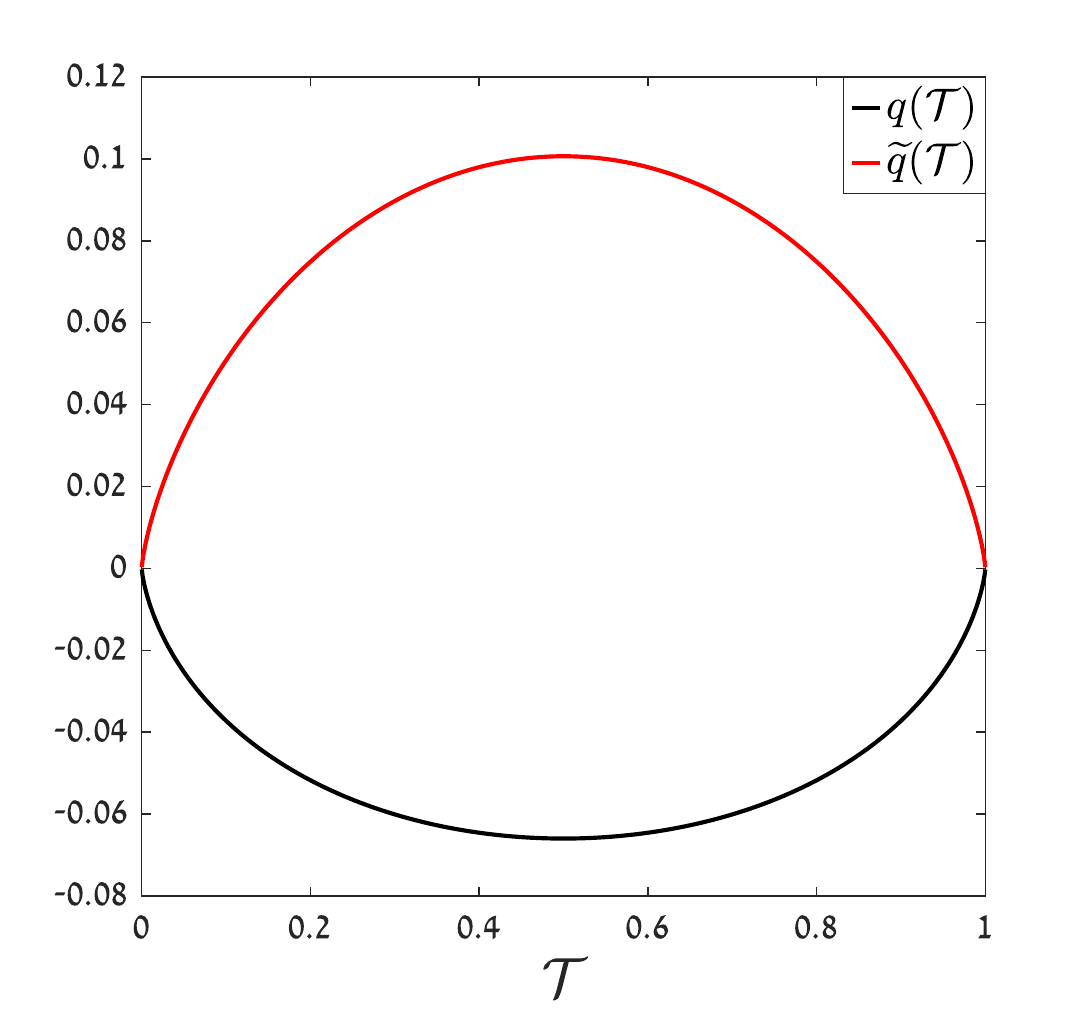}
\par\end{centering}
\caption{\label{fig:q-functions}The functions $q\!\left({\cal T}\right)$
and $\widetilde{q}\!\left({\cal T}\right)$, given in Eqs.~(\ref{eq:MI-LogCoefficient-Overlap})
and (\ref{eq:MI-LogCoefficient-NoOverlap}), respectively.}

\end{figure}

Additionally, for $0\le{\cal T}\le1$ and ${\cal R}=1-{\cal T}$,
we define
\begin{align}
\widetilde{Q}_{n}\!\left({\cal T}\right) & =Q_{n}\!\left({\cal T}\right)+Q_{n}\!\left({\cal R}\right)+\frac{n}{2\pi^{2}}\int_{{\cal R}}^{{\cal T}}dx\frac{x^{n-1}-\left(1-x\right)^{n-1}}{x^{n}+\left(1-x\right)^{n}}\ln\!\left|\frac{{\cal R}-x}{{\cal T}-x}\right|\nonumber \\
 & =-\frac{n}{12}+\int_{0}^{1}\frac{dx}{2\pi^{2}x}\left\{ \ln\!\left[\left(1+{\cal T}x\right)^{n}+\left({\cal R}x\right)^{n}\right]+\ln\!\left[\left(1+{\cal R}x\right)^{n}+\left({\cal T}x\right)^{n}\right]\right\} \nonumber \\
 & +\int_{0}^{1}\frac{dx}{2\pi^{2}x}\left\{ \ln\!\left[\frac{\left(x+{\cal T}\right)^{n}+{\cal R}^{n}}{\left({\cal T}+{\cal R}x\right)^{n}+\left({\cal R}+{\cal T}x\right)^{n}}\right]+\ln\!\left[\frac{\left(x+{\cal R}\right)^{n}+{\cal T}^{n}}{\left({\cal T}+{\cal R}x\right)^{n}+\left({\cal R}+{\cal T}x\right)^{n}}\right]\right\} ,\label{eq:Log-scaling-kernel-no-overlap}
\end{align}
as well as the functions
\begin{align}
q\!\left({\cal T}\right) & =\frac{1}{24}-\int_{0}^{1}\frac{dx}{2\pi^{2}x}\cdot\frac{\left(1+{\cal R}x\right)\ln\!\left(1+{\cal R}x\right)+\left(1+{\cal T}x\right)\ln\!\left(1+{\cal T}x\right)}{1+x}\nonumber \\
 & +\int_{0}^{1}\frac{dx}{2\pi^{2}x}\left[{\cal T}\ln{\cal T}+{\cal R}\ln{\cal R}-\frac{\left({\cal R}+x\right)\ln\!\left({\cal R}+x\right)+\left({\cal T}+x\right)\ln\!\left({\cal T}+x\right)}{1+x}\right],\label{eq:MI-LogCoefficient-Overlap}
\end{align}
and
\begin{equation}
\widetilde{q}\!\left({\cal T}\right)=q\!\left({\cal T}\right)+\frac{1}{12}+\int_{0}^{1}\frac{dx}{\pi^{2}x}\left[\frac{\left({\cal R}+{\cal T}x\right)\ln\!\left({\cal R}+{\cal T}x\right)+\left({\cal T}+{\cal R}x\right)\ln\!\left({\cal T}+{\cal R}x\right)}{1+x}-{\cal T}\ln{\cal T}-{\cal R}\ln{\cal R}\right].\label{eq:MI-LogCoefficient-NoOverlap}
\end{equation}
The functions $q\!\left({\cal T}\right)$ and $\widetilde{q}\!\left({\cal T}\right)$
will appear in the analytical expression for the logarithmic term
of the MI. To make their dependence on ${\cal T}$ more apparent,
we plot them in Fig.~\ref{fig:q-functions}, where one may observe
that they both vanish for ${\cal T}=0,1$, while for $0<{\cal T}<1$
we have $q\!\left({\cal T}\right)<0$ and $\widetilde{q}\!\left({\cal T}\right)>0$.

\section{Asymptotics beyond volume-law entanglement\label{sec:Asymptotics-beyond-volume-law}}

\subsection{Main results\label{subsec:Main-results}}

The new results that we present in this work apply to the long-range
limit: $d_{i}/\ell_{i}\to\infty$, with $d_{{\scriptscriptstyle L}}-d_{{\scriptscriptstyle R}}$
kept fixed. The volume-law asymptotics that have been previously derived
(see Subsec.~\ref{subsec:Recap}) are independent of the absolute
distances of the subsystems from the impurity, but subleading corrections
are in general sensitive to these distances. We thus focus on the
long-range limit so that our results capture the nature of the unique
long-range correlations produced in the steady state.

Let $m_{1}\le m_{2}\le m_{3}\le m_{4}$ denote the lengths $d_{{\scriptscriptstyle L}}$,
$\ell_{{\scriptscriptstyle L}}+d_{{\scriptscriptstyle L}}$, $d_{{\scriptscriptstyle R}}$
and $\ell_{{\scriptscriptstyle R}}+d_{{\scriptscriptstyle R}}$ in
ascending order. Our analysis shows that, in the long-range limit,
the asymptotics of the R\'enyi MI between $A_{{\scriptscriptstyle L}}$
and $A_{{\scriptscriptstyle R}}$ is given by
\begin{align}
{\cal I}_{A_{L}:A_{R}}^{\left(n\right)} & \sim\frac{\ell_{{\rm mirror}}}{1-n}\int_{k_{-}}^{k_{+}}\frac{dk}{\pi}\ln\!\left[\left({\cal T}\!\left(k\right)\right)^{n}+\left({\cal R}\!\left(k\right)\right)^{n}\right]\nonumber \\
 & +\frac{1}{2\left(1-n\right)}\left[{\cal G}_{n}^{\left({\rm log}\right)}\!\left(k_{{\scriptscriptstyle F,L}},\ell_{{\scriptscriptstyle L}},\ell_{{\scriptscriptstyle R}},d_{{\scriptscriptstyle L}}-d_{{\scriptscriptstyle R}}\right)+{\cal G}_{n}^{\left({\rm log}\right)}\!\left(k_{{\scriptscriptstyle F,R}},\ell_{{\scriptscriptstyle L}},\ell_{{\scriptscriptstyle R}},d_{{\scriptscriptstyle L}}-d_{{\scriptscriptstyle R}}\right)\right],\label{eq:Renyi-MI-full-asymptotics}
\end{align}
where we defined
\begin{align}
{\cal G}_{n}^{\left({\rm log}\right)}\!\left(k_{{\scriptscriptstyle F}},\ell_{{\scriptscriptstyle L}},\ell_{{\scriptscriptstyle R}},d_{{\scriptscriptstyle L}}-d_{{\scriptscriptstyle R}}\right) & =\left(Q_{n}\!\left({\cal T}\!\left(k_{{\scriptscriptstyle F}}\right)\right)+Q_{n}\!\left({\cal R}\!\left(k_{{\scriptscriptstyle F}}\right)\right)-\frac{1}{12}\left(\frac{1}{n}-n\right)\right)\ln\!\left|\frac{\left(m_{3}-m_{1}\right)\left(m_{4}-m_{2}\right)}{\left(\ell_{{\scriptscriptstyle L}}+d_{{\scriptscriptstyle L}}-\ell_{{\scriptscriptstyle R}}-d_{{\scriptscriptstyle R}}\right)\left(d_{{\scriptscriptstyle L}}-d_{{\scriptscriptstyle R}}\right)}\right|\nonumber \\
 & +\widetilde{Q}_{n}\!\left({\cal T}\!\left(k_{{\scriptscriptstyle F}}\right)\right)\ln\!\left|\frac{\left(m_{3}-m_{1}\right)\left(m_{4}-m_{2}\right)}{\left(\ell_{{\scriptscriptstyle R}}+d_{{\scriptscriptstyle R}}-d_{{\scriptscriptstyle L}}\right)\left(\ell_{{\scriptscriptstyle L}}+d_{{\scriptscriptstyle L}}-d_{{\scriptscriptstyle R}}\right)}\right|.\label{eq:Renyi-MI-log-term}
\end{align}
Eq.~(\ref{eq:Renyi-MI-full-asymptotics}) may be taken to the limit
$n\to1$ to obtain the von Neumann MI. This yields
\begin{align}
{\cal I}_{A_{L}:A_{R}} & \sim\ell_{{\rm mirror}}\int_{k_{-}}^{k_{+}}\frac{dk}{\pi}\left[-{\cal T}\!\left(k\right)\ln{\cal T}\!\left(k\right)-{\cal R}\!\left(k\right)\ln{\cal R}\!\left(k\right)\right]\nonumber \\
 & +\frac{1}{2}\left[g^{\left({\rm log}\right)}\!\left(k_{{\scriptscriptstyle F,L}},\ell_{{\scriptscriptstyle L}},\ell_{{\scriptscriptstyle R}},d_{{\scriptscriptstyle L}}-d_{{\scriptscriptstyle R}}\right)+g^{\left({\rm log}\right)}\!\left(k_{{\scriptscriptstyle F,R}},\ell_{{\scriptscriptstyle L}},\ell_{{\scriptscriptstyle R}},d_{{\scriptscriptstyle L}}-d_{{\scriptscriptstyle R}}\right)\right].\label{eq:Asymptotics_of_von_Neumann_MI}
\end{align}
Here we defined the following notation to represent the contribution
of each Fermi momentum to the logarithmic term of the MI:
\begin{align}
g^{\left({\rm log}\right)}\!\left(k_{{\scriptscriptstyle F}},\ell_{{\scriptscriptstyle L}},\ell_{{\scriptscriptstyle R}},d_{{\scriptscriptstyle L}}-d_{{\scriptscriptstyle R}}\right) & =q\!\left({\cal T}\!\left(k_{{\scriptscriptstyle F}}\right)\right)\ln\!\left|\frac{\left(m_{3}-m_{1}\right)\left(m_{4}-m_{2}\right)}{\left(\ell_{{\scriptscriptstyle L}}+d_{{\scriptscriptstyle L}}-\ell_{{\scriptscriptstyle R}}-d_{{\scriptscriptstyle R}}\right)\left(d_{{\scriptscriptstyle L}}-d_{{\scriptscriptstyle R}}\right)}\right|\nonumber \\
 & +\widetilde{q}\!\left({\cal T}\!\left(k_{{\scriptscriptstyle F}}\right)\right)\ln\!\left|\frac{\left(m_{3}-m_{1}\right)\left(m_{4}-m_{2}\right)}{\left(\ell_{{\scriptscriptstyle R}}+d_{{\scriptscriptstyle R}}-d_{{\scriptscriptstyle L}}\right)\left(\ell_{{\scriptscriptstyle L}}+d_{{\scriptscriptstyle L}}-d_{{\scriptscriptstyle R}}\right)}\right|;\label{eq:MI-LogTerm}
\end{align}
as can be seen in Eq.~(\ref{eq:Asymptotics_of_von_Neumann_MI}),
these contributions are weighted equally. It is noteworthy that the
two terms appearing in Eq.~(\ref{eq:MI-LogTerm}) have distinct signs.
Suppose that $0<{\cal T}\!\left(k_{{\scriptscriptstyle F}}\right)<1$,
then the first summand in Eq.~(\ref{eq:MI-LogTerm}) vanishes if
$\ell_{{\rm mirror}}=0$, and is negative for $\ell_{{\rm mirror}}>0$,
meaning that it decreases the correlation between $A_{{\scriptscriptstyle L}}$
and $A_{{\scriptscriptstyle R}}$ when they have nonzero mirror-image
overlap. On the other hand, the second summand vanishes if either
$\Delta\ell_{{\scriptscriptstyle L}}=0$ or $\Delta\ell_{{\scriptscriptstyle R}}=0$,
and is positive otherwise; this implies that this term increases the
correlation between $A_{{\scriptscriptstyle L}}$ and $A_{{\scriptscriptstyle R}}$
unless their mirror-image overlap is maximal (i.e., the mirror image
of one of them is contained in the other).

We stress that Eqs.~(\ref{eq:Renyi-MI-full-asymptotics}) and (\ref{eq:Asymptotics_of_von_Neumann_MI})
can be applied also to degenerate cases where differences appearing
inside the logarithms vanish; the apparent conundrum is solved simply
by omitting the problematic difference (as will be justified by the
derivation). For example, if $A_{{\scriptscriptstyle R}}$ has no
overlap with the mirror image of $A_{{\scriptscriptstyle L}}$ but
they exactly touch each other, such that $\ell_{{\scriptscriptstyle R}}+d_{{\scriptscriptstyle R}}=d_{{\scriptscriptstyle L}}$,
then in Eq.~(\ref{eq:Renyi-MI-full-asymptotics}) both the linear
term and the first logarithmic term from Eq.~(\ref{eq:Renyi-MI-log-term})
vanish, while in the second logarithmic term from Eq.~(\ref{eq:Renyi-MI-log-term})
we drop the $\left(\ell_{{\scriptscriptstyle R}}+d_{{\scriptscriptstyle R}}-d_{{\scriptscriptstyle L}}\right)$
term appearing in the denominator. Eq.~(\ref{eq:Renyi-MI-full-asymptotics})
then reads
\begin{align}
{\cal I}_{A_{L}:A_{R}}^{\left(n\right)} & \sim\frac{\widetilde{Q}_{n}\!\left({\cal T}\!\left(k_{{\scriptscriptstyle F,L}}\right)\right)+\widetilde{Q}_{n}\!\left({\cal T}\!\left(k_{{\scriptscriptstyle F,R}}\right)\right)}{2\left(1-n\right)}\ln\!\left(\frac{\ell_{{\scriptscriptstyle L}}\ell_{{\scriptscriptstyle R}}}{\ell_{{\scriptscriptstyle L}}+\ell_{{\scriptscriptstyle R}}}\right).
\end{align}

The recipe that we will present for the calculation of the MI asymptotics
can be also applied to the calculation of the fermionic negativity
asymptotics, up to the logarithmic order. In Subsec.~\ref{subsec:Derivation-of-negativity}
we elaborate on how this extension of the recipe can be done in principle,
though its complete execution is more cumbersome than in the case
of the MI calculation. For this reason we present the full result
for the negativity asymptotics only in the symmetric case with $\ell_{{\scriptscriptstyle L}}=\ell_{{\scriptscriptstyle R}}\equiv\ell$
and $d_{{\scriptscriptstyle L}}=d_{{\scriptscriptstyle R}}$ (again,
the limit $d_{i}/\ell_{i}\to\infty$ is taken). The R\'enyi negativities
are then given by
\begin{align}
{\cal E}_{n} & \sim\ell\int_{k_{-}}^{k_{+}}\frac{dk}{\pi}\ln\!\left[\left({\cal T}\!\left(k\right)\right)^{n/2}+\left({\cal R}\!\left(k\right)\right)^{n/2}\right]-\frac{n}{4}\ln\ell\nonumber \\
 & +\ln\ell\left[Q_{n/2}\!\left({\cal T}\!\left(k_{{\scriptscriptstyle F,L}}\right)\right)+Q_{n/2}\!\left({\cal R}\!\left(k_{{\scriptscriptstyle F,L}}\right)\right)+Q_{n/2}\!\left({\cal T}\!\left(k_{{\scriptscriptstyle F,R}}\right)\right)+Q_{n/2}\!\left({\cal R}\!\left(k_{{\scriptscriptstyle F,R}}\right)\right)\right],\label{eq:Renyi-negativity-full-asymptotics-symmetric}
\end{align}
and the fermionic negativity is accordingly given by their $n\to1$
limit, namely

\begin{align}
{\cal E} & \sim\ell\int_{k_{-}}^{k_{+}}\frac{dk}{\pi}\ln\!\left[\left({\cal T}\!\left(k\right)\right)^{1/2}+\left({\cal R}\!\left(k\right)\right)^{1/2}\right]-\frac{1}{4}\ln\ell\nonumber \\
 & +\ln\ell\left[Q_{1/2}\!\left({\cal T}\!\left(k_{{\scriptscriptstyle F,L}}\right)\right)+Q_{1/2}\!\left({\cal R}\!\left(k_{{\scriptscriptstyle F,L}}\right)\right)+Q_{1/2}\!\left({\cal T}\!\left(k_{{\scriptscriptstyle F,R}}\right)\right)+Q_{1/2}\!\left({\cal R}\!\left(k_{{\scriptscriptstyle F,R}}\right)\right)\right].\label{eq:Negativity-full-asymptotics-symmetric}
\end{align}

We emphasize that the R\'enyi MI, the MI and the negativity all vanish
either in the presence of a trivial impurity (with ${\cal T}\!\left(k\right)\in\left\{ 0,1\right\} $
for all $k$) or in the absence of a bias. In Subsec.~\ref{subsec:Recap}
we noted that this property holds for the leading-order extensive
terms of these quantities, which were derived in Ref.~\citep{fraenkel2022extensive};
in Subsec.~\ref{subsec:Correlation-matrix} we show that it is true
at all orders, since in either of the two cases the correlations between
$A_{{\scriptscriptstyle L}}$ and $A_{{\scriptscriptstyle R}}$ vanish
for $d_{i}/\ell_{i}\to\infty$ (the R\'enyi negativities do not vanish,
in congruence with them not being proper correlation measures). In
particular, this of course implies that the logarithmic terms in Eqs.~(\ref{eq:Renyi-MI-full-asymptotics}),
(\ref{eq:Asymptotics_of_von_Neumann_MI}), and (\ref{eq:Negativity-full-asymptotics-symmetric})
should vanish in both cases. In the case of a trivial impurity this
is directly seen by substituting ${\cal T}=0$ or ${\cal T}=1$ into
the logarithmic terms (in Eq.~(\ref{eq:Renyi-MI-full-asymptotics})
we use the fact that $\widetilde{Q}_{n}\!\left(1\right)=\widetilde{Q}_{n}\!\left(0\right)=0$),
while these terms do not vanish if one simply substitutes $k_{{\scriptscriptstyle F,L}}=k_{{\scriptscriptstyle F,R}}$
into the equations. Like in the case of single-interval entropies,
this discrepancy occurs due to non-commuting limits (see the discussion
following Eq.~(\ref{eq:Renyi-entropies-single-intervals})), and
we point it out in order to stress that our analytical expressions
rely on the assumption of a finite bias, and one could be misled by
naively taking their limits to the equilibrium scenario. Since for
all the quantities that we consider here the extensive term tends
to zero near the no-bias limit while the logarithmic term does not,
it is clear that the logarithmic term becomes especially significant
when the bias is small but finite.

\subsection{Comparison to numerics\label{subsec:Comparison-to-numerics}}

\begin{figure}
\begin{centering}
\includegraphics[viewport=40bp 90bp 1168bp 750bp,clip,width=1\columnwidth]{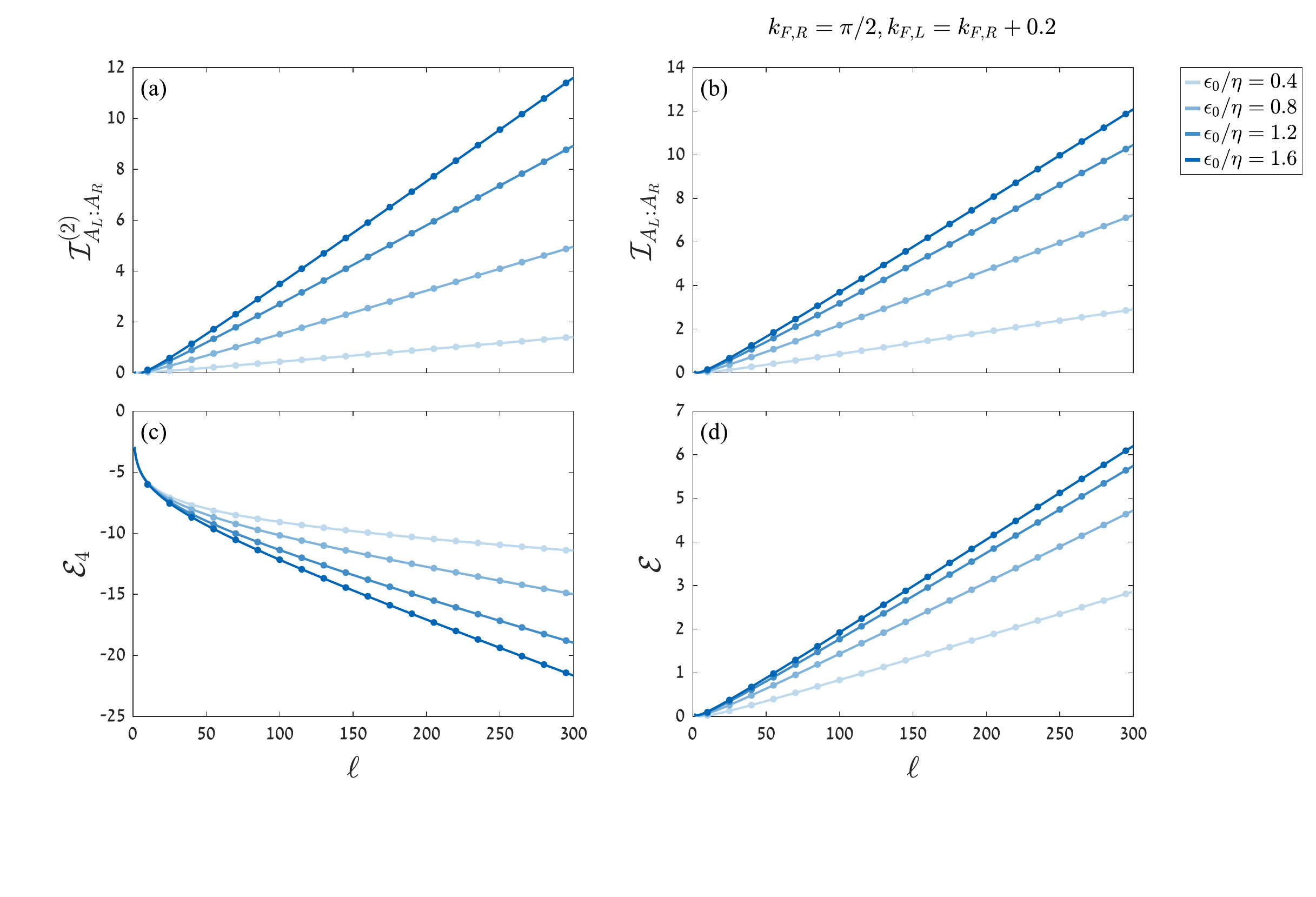}
\par\end{centering}
\caption{\label{fig:Scaling-of-measures-symmetric}Scaling of (a) the 2-R\'enyi
mutual information, (b) the mutual information, (c) the 4-R\'enyi
negativity and (d) the fermionic negativity between subsystems $A_{{\scriptscriptstyle L}}$
and $A_{{\scriptscriptstyle R}}$ in the single-site impurity model,
for a symmetric subsystem configuration with $\ell_{{\scriptscriptstyle L}}=\ell_{{\scriptscriptstyle R}}\equiv\ell$
and $d_{{\scriptscriptstyle L}}=d_{{\scriptscriptstyle R}}$. The
long-range limit $d_{i}/\ell_{i}\to\infty$ is taken, and the results
are plotted for various values of the impurity energy $\epsilon_{0}$,
while setting the Fermi momenta at $k_{{\scriptscriptstyle F,R}}=\pi/2$
and $k_{{\scriptscriptstyle F,L}}=k_{{\scriptscriptstyle F,R}}+0.2$.
In all the panels, continuous lines designate the analytical results
(Eqs.~(\ref{eq:Renyi-MI-full-asymptotics}), (\ref{eq:Asymptotics_of_von_Neumann_MI}),
(\ref{eq:Renyi-negativity-full-asymptotics-symmetric}) and (\ref{eq:Negativity-full-asymptotics-symmetric}),
respectively; an additive constant-in-$\ell$ term constitutes the
only fitting parameter), while dots represent numerical results (the
computation of which is discussed in Subsec.~\ref{subsec:Correlation-matrix}).}
\end{figure}

In order to corroborate our asymptotic analysis, we checked the analytical
results reported in Subsec.~\ref{subsec:Main-results} against numerical
computations. The correlation measures discussed in this paper can
be computed numerically through exact diagonalization of two-point
correlation matrices, as further explained in Subsec.~\ref{subsec:Correlation-matrix};
there we also provide the explicit expression for the two-point correlation
function in the limit $d_{i}/\ell_{i}\to\infty$ (see Eq.~(\ref{eq:Correlation-function-large-distance})).

As a specific example for a noninteracting impurity, we choose a single-site
impurity with a nonzero on-site energy. That is, in the Hamiltonian
of Eq.~(\ref{eq:Model-Hamiltonian}) we set $m_{0}=0$ and ${\cal H}_{{\rm scat}}=\epsilon_{0}c_{0}^{\dagger}c_{0}$,
with $\epsilon_{0}$ being the impurity energy. The corresponding
transmission probability is given by
\begin{equation}
{\cal T}\!\left(k\right)=\frac{\sin^{2}\!k}{\sin^{2}\!k+\left(\epsilon_{0}/2\eta\right)^{2}}.\label{eq:Single-impurity-transmission}
\end{equation}

In Fig.~\ref{fig:Scaling-of-measures-symmetric} we show a comparison
between our analytical results and numerics, for the symmetric configuration
where the two subsystems of interest are of equal length and within
an equal distance from the impurity. We plot the scaling with subsystem
length of the MI, negativity, 2-R\'enyi MI and 4-R\'enyi negativity,
where the analytical results include the exact expressions for the
linear and logarithmic terms, along with an additive constant correction
constituting the only fitting parameter. The excellent agreement of
our results with the numerics is clearly evident.

\begin{figure}
\begin{centering}
\includegraphics[viewport=100bp 0bp 1350bp 465bp,clip,width=1\columnwidth]{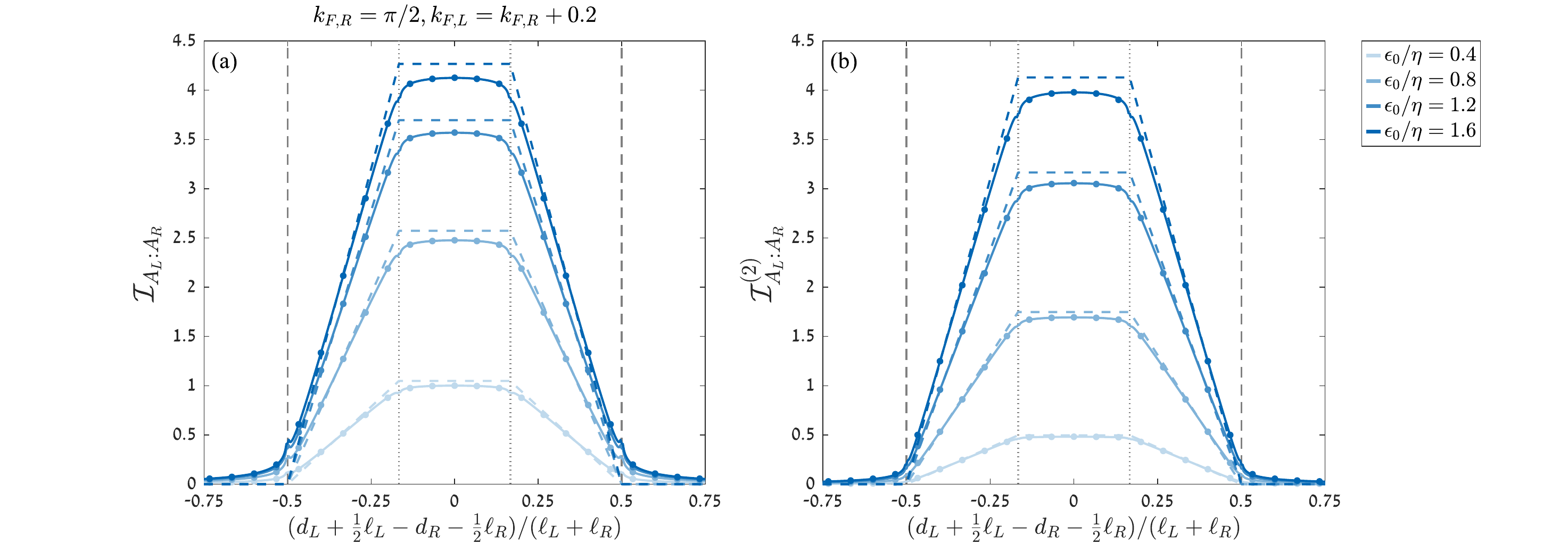}
\par\end{centering}
\caption{\label{fig:Scaling-of-MI-asymmetric}(a) Scaling of the mutual information
between subsystems $A_{{\scriptscriptstyle L}}$ and $A_{{\scriptscriptstyle R}}$
in the single-site impurity model, for fixed subsystem lengths ($\ell_{{\scriptscriptstyle L}}=100$
and $\ell_{{\scriptscriptstyle R}}=200$) and varying $d_{{\scriptscriptstyle L}}-d_{{\scriptscriptstyle R}}$.
The long-range limit $d_{i}/\ell_{i}\to\infty$ is taken, and the
results are plotted for various values of the impurity energy $\epsilon_{0}$,
while setting the Fermi momenta at $k_{{\scriptscriptstyle F,R}}=\pi/2$
and $k_{{\scriptscriptstyle F,L}}=k_{{\scriptscriptstyle F,R}}+0.2$.
Continuous lines represent the analytical result of Eq.~(\ref{eq:Asymptotics_of_von_Neumann_MI}),
dashed blue lines represent the analytical result including only the
leading-order volume-law term, and dots represent the numerical results
(the computation of which is discussed in Subsec.~\ref{subsec:Correlation-matrix}).
Employing the notation $\bar{A}_{{\scriptscriptstyle L}}=\left\{ m|-\!m\in A_{{\scriptscriptstyle L}}\right\} $
for the mirror image of $A_{{\scriptscriptstyle L}}$, dotted vertical
lines designate the range of values of $d_{{\scriptscriptstyle L}}-d_{{\scriptscriptstyle R}}$
in which $\bar{A}_{{\scriptscriptstyle L}}\subset A_{{\scriptscriptstyle R}}$,
while dashed vertical lines mark the range in which $\bar{A}_{{\scriptscriptstyle L}}\cap A_{{\scriptscriptstyle R}}\protect\neq\phi$.
Panel (b) shows a similar analysis for the 2-R\'enyi mutual information,
where the analytical results were computed using Eq.~(\ref{eq:Renyi-MI-full-asymptotics}).}
\end{figure}

This importance of the logarithmic correction, which is the refinement
provided by the current work relative to the results reported in Ref.~\citep{fraenkel2022extensive},
becomes especially conspicuous when we compare the analytical results
and the numerics for a scenario where we fix the lengths of the two
subsystems and vary the position of one of them. In Fig.~\ref{fig:Scaling-of-MI-asymmetric}
we plot for such a scenario the 2-R\'enyi MI and the MI, for which
we explicitly derived the logarithmic terms for a general subsystem
configuration, given in Eqs.~(\ref{eq:Renyi-MI-full-asymptotics})
and (\ref{eq:Asymptotics_of_von_Neumann_MI}), respectively. As expected
from the form of the leading-order terms appearing in Eqs.~(\ref{eq:Renyi-MI-full-asymptotics})
and (\ref{eq:Asymptotics_of_von_Neumann_MI}), these measures reach
a maximal value for a maximal mirror-image overlap between the two
subsystems. However, the corrections beyond the leading order are
clearly substantial: by plotting the results of Eqs.~(\ref{eq:Renyi-MI-full-asymptotics})
and (\ref{eq:Asymptotics_of_von_Neumann_MI}) both with the logarithmic
terms and without them, we observe that their inclusion leads to a
much more accurate agreement with the numerics.

Finally, we should also emphasize that the derivation of the logarithmic
terms in Eqs.~(\ref{eq:Renyi-MI-full-asymptotics}) and (\ref{eq:Asymptotics_of_von_Neumann_MI})
relies on the study of a simplified version of the nonequilibrium
steady state, where the scattering matrix associated with the impurity
is assumed to be independent of the energy of the incoming particle.
The extension of the results to energy-dependent scattering does not
rely on a direct computation, but is rather conjectural; it relies
on an analogy with the exact form of the logarithmic term for a symmetric
subsystem configuration, for which a direct computation is indeed
possible (see more details in Subsec.~\ref{subsec:Derivation-of-MI}).
Since our choice of an impurity model gives rise to an energy-dependent
transmission probability (Eq.~(\ref{eq:Single-impurity-transmission})),
the comparison shown in Fig.~\ref{fig:Scaling-of-MI-asymmetric}
is a nontrivial verification of this conjecture. To further confirm
the apparent agreement between numerical and analytical results in
Fig.~\ref{fig:Scaling-of-MI-asymmetric}, in Appendix \ref{sec:Numerics-comparison-appendix}
we plot the differences between them and briefly discuss the behavior
of the deviation.

\section{Derivation of the analytical results\label{sec:Derivation-of-results}}

This section describes in the detail the derivation of the results
reported in Sec.~\ref{sec:Asymptotics-beyond-volume-law}. First,
in Subsec.~\ref{subsec:Correlation-matrix} we discuss the two-point
correlation structure of the steady state and its mathematical relations
with the correlation measures of interest. Based on these relations,
in Subsec.~\ref{subsec:Derivation-of-MI} we derive the analytical
expression for the R\'enyi MI (given in Eq.~(\ref{eq:Renyi-MI-full-asymptotics})),
and in Subsec.~\ref{subsec:Derivation-of-negativity} we show how
the same method can be employed for the computation of R\'enyi negativities.

\subsection{The correlation matrix and its long-range limit\label{subsec:Correlation-matrix}}

Given a Gaussian state of the system, like the steady state considered
in this work, subsystem correlation measures are fully determined
by the restricted two-point correlation matrix $C_{X}$, with entries
$\left(C_{X}\right)_{jm}=\left\langle c_{j}^{\dagger}c_{m}\right\rangle $
($j,m\in X$). In particular, the R\'enyi entropy of any subsystem
$X$ may be expressed as \citep{Peschel_2003}
\begin{equation}
S_{X}^{\left(n\right)}=\frac{1}{1-n}{\rm Tr}\ln\!\left[\left(C_{X}\right)^{n}+\left(\mathbb{I}-C_{X}\right)^{n}\right].\label{eq:Entropy_from_correlations}
\end{equation}
In a similar fashion, the R\'enyi negativity ${\cal E}_{n}$ between
two subsystems $X_{1}$ and $X_{2}$ within a Gaussian state can be
generally written as \citep{PhysRevB.95.165101,PhysRevB.97.165123,Shapourian_2019}
\begin{equation}
{\cal E}_{n}={\rm Tr}\ln\!\left[\left(C_{\Xi}\right)^{n/2}+\left(\mathbb{I}-C_{\Xi}\right)^{n/2}\right]+\frac{n}{2}{\rm Tr}\ln\!\left[\left(C_{X_{1}\cup X_{2}}\right)^{2}+\left(\mathbb{I}-C_{X_{1}\cup X_{2}}\right)^{2}\right].\label{eq:Negativity_from_correlations}
\end{equation}
Here $C_{\Xi}$ is a transformed correlation matrix, defined as
\begin{equation}
C_{\Xi}=\frac{1}{2}\left[\mathbb{I}-\left(\mathbb{I}+\Gamma_{+}\Gamma_{-}\right)^{-1}\left(\Gamma_{+}+\Gamma_{-}\right)\right],
\end{equation}
with 
\begin{equation}
\Gamma_{\pm}=\left(\begin{array}{cc}
\pm i\mathbb{I}_{\left|X_{1}\right|} & 0\\
0 & \mathbb{I}_{\left|X_{2}\right|}
\end{array}\right)\left(\mathbb{I}-2C_{X_{1}\cup X_{2}}\right)\left(\begin{array}{cc}
\pm i\mathbb{I}_{\left|X_{1}\right|} & 0\\
0 & \mathbb{I}_{\left|X_{2}\right|}
\end{array}\right),
\end{equation}
assuming the entries of $C_{X_{1}\cup X_{2}}$ are ordered such that
its first $\left|X_{1}\right|$ indices ($\left|X_{i}\right|$ being
the size of $X_{i}$) correspond to the sites of $X_{1}$. In Ref.~\citep{fraenkel2022extensive}
we showed that Eq.~(\ref{eq:Negativity_from_correlations}) is in
fact equivalent to
\begin{equation}
{\cal E}_{n}={\rm Tr}\ln\!\left[\prod_{\gamma=-\frac{n-1}{2}}^{\frac{n-1}{2}}\left(\mathbb{I}-C_{\gamma}\right)\right],\label{eq:Negativity_from_correlations_variation}
\end{equation}
with $C_{\gamma}$ defined for $\gamma=-\frac{n-1}{2},-\frac{n-3}{2},\ldots,\frac{n-1}{2}$
as 
\begin{equation}
C_{\gamma}=\left(\begin{array}{cc}
\left(1-e^{\frac{2\pi i\gamma}{n}}\right)\mathbb{I}_{\left|X_{1}\right|} & 0\\
0 & \left(1+e^{\frac{-2\pi i\gamma}{n}}\right)\mathbb{I}_{\left|X_{2}\right|}
\end{array}\right)C_{X_{1}\cup X_{2}}.
\end{equation}
Eqs.~(\ref{eq:Entropy_from_correlations}), (\ref{eq:Negativity_from_correlations})
and (\ref{eq:Negativity_from_correlations_variation}) are general
formulae (applicable to any Gaussian state) that constitute the basis
for the computation of all analytical and numerical results reported
in this paper.

For the particular state considered in this paper, the two-point correlation
function for sites $j,m$ outside the scattering region (such that
$\left|j\right|,\left|m\right|>m_{0}$) can be extracted from the
relation
\begin{equation}
\left\langle c_{j}^{\dagger}c_{m}\right\rangle =\int_{-k_{F,R}}^{k_{F,L}}\frac{dk}{2\pi}\,\left\langle k|j\right\rangle \left\langle m|k\right\rangle ,\label{eq:Correlation_function_integral}
\end{equation}
using the scattering state wavefunctions given in Eqs.~(\ref{eq:Left-scattering-states})--(\ref{eq:Right-scattering-states}).
In Ref.~\citep{fraenkel2022extensive} we observed that in the long-range
limit where $d_{i}/\ell_{i}\to\infty$ with $d_{{\scriptscriptstyle L}}-d_{{\scriptscriptstyle R}}$
kept fixed, the explicit expression for the correlation function in
Eq.~(\ref{eq:Correlation_function_integral}) is simplified, as certain
contributions vanish in this limit. Accordingly, for $d_{i}/\ell_{i}\gg1$
we may use the approximation

\begin{equation}
\left\langle c_{j}^{\dagger}c_{m}\right\rangle \approx\begin{cases}
\int_{-k_{F,R}}^{k_{F,R}}\frac{dk}{2\pi}\,e^{-i\left(j-m\right)k}+\int_{k_{F,R}}^{k_{F,L}}\frac{dk}{2\pi}\,{\cal T}\!\left(k\right)e^{-i\left(j-m\right)k} & j,m\in A_{{\scriptscriptstyle R}},\\
\int_{-k_{F,L}}^{k_{F,L}}\frac{dk}{2\pi}\,e^{-i\left(j-m\right)k}+\int_{k_{F,L}}^{k_{F,R}}\frac{dk}{2\pi}\,{\cal T}\!\left(k\right)e^{i\left(j-m\right)k} & j,m\in A_{{\scriptscriptstyle L}},\\
\int_{k_{F,R}}^{k_{F,L}}\frac{dk}{2\pi}\,t_{{\scriptscriptstyle L}}^{*}\!\left(k\right)r_{{\scriptscriptstyle L}}\!\left(k\right)e^{-i\left(j+m\right)k} & m\in A_{{\scriptscriptstyle L}}\text{ and }j\in A_{{\scriptscriptstyle R}},\\
\int_{k_{F,L}}^{k_{F,R}}\frac{dk}{2\pi}\,t_{{\scriptscriptstyle R}}^{*}\!\left(k\right)r_{{\scriptscriptstyle R}}\!\left(k\right)e^{i\left(j+m\right)k} & j\in A_{{\scriptscriptstyle L}}\text{ and }m\in A_{{\scriptscriptstyle R}}.
\end{cases}\label{eq:Correlation-function-large-distance}
\end{equation}

Since the long-range limit is the focus of this paper, Eq.~(\ref{eq:Correlation-function-large-distance})
was the expression used for the correlation matrix entries when numerical
results were computed using Eqs.~(\ref{eq:Entropy_from_correlations})--(\ref{eq:Negativity_from_correlations}).
As we explain in detail in the following subsections, Eq.~(\ref{eq:Correlation-function-large-distance})
was also used as the starting point for some parts of the analytical
derivation, whereas in other parts the long-range limit was taken
only in a later stage of the calculation.

We may point out that Eq.~(\ref{eq:Correlation-function-large-distance})
entails that, in the long-range limit, the two-point correlation function
between a site in $A_{{\scriptscriptstyle L}}$ and another in $A_{{\scriptscriptstyle R}}$
vanishes either when the impurity is trivial or when $k_{{\scriptscriptstyle F,L}}=k_{{\scriptscriptstyle F,R}}$.
Then, it is straightforward to see through Eqs.~(\ref{eq:Entropy_from_correlations})
and (\ref{eq:Negativity_from_correlations}) that in either of these
two cases the R\'enyi MI vanishes identically and the R\'enyi negativity
satisfies ${\cal E}_{n}=\left(1-n\right)S_{A}^{\left(n\right)}$,
meaning that both the (von Neumann) MI and the negativity vanish.

\subsection{Derivation of the mutual information asymptotics\label{subsec:Derivation-of-MI}}

In this subsection we discuss the derivation of Eq.~(\ref{eq:Renyi-MI-full-asymptotics}),
which reports the asymptotics of the R\'enyi MI in the long-range
limit. As $S_{A_{L}}^{\left(n\right)}$ and $S_{A_{R}}^{\left(n\right)}$
are known exactly up to the logarithmic order from Ref.~\citep{10.21468/SciPostPhys.11.4.085}
(recall Eq.~(\ref{eq:Renyi-entropies-single-intervals})), what remains
to be done is the computation of $S_{A}^{\left(n\right)}$, the R\'enyi
entropy of the union of the two intervals. First, in Subec.~\ref{subsec:MI-derivation-Symmetric-configuration},
we perform the computation in the case where the two intervals are
of equal length, $\ell_{{\scriptscriptstyle L}}=\ell_{{\scriptscriptstyle R}}$,
a choice which allows to treat the problem exactly for certain regimes
of the value of $\left|d_{{\scriptscriptstyle L}}-d_{{\scriptscriptstyle R}}\right|$.
Then, in the remainder of this subsection, we tackle the more general
calculation (including $\ell_{{\scriptscriptstyle L}}\neq\ell_{{\scriptscriptstyle R}}$,
and for any choice of $d_{{\scriptscriptstyle L}}-d_{{\scriptscriptstyle R}}$)
by studying a simplified version of the steady state, and by relying
on insights obtained from the exact solution featured in Subsec.~\ref{subsec:MI-derivation-Symmetric-configuration}.

\subsubsection{Special case: Equal-length intervals\label{subsec:MI-derivation-Symmetric-configuration}}

We thus begin by focusing on the case where $A_{{\scriptscriptstyle L}}$
and $A_{{\scriptscriptstyle R}}$ are of equal length, $\ell_{{\scriptscriptstyle L}}=\ell_{{\scriptscriptstyle R}}\equiv\ell$.
The restricted two-point correlation matrix $C_{A}$ can then be written
in terms of $2\times2$ blocks, 

\begin{equation}
\Phi_{jm}=\left(\begin{array}{cc}
\left\langle c_{m_{0}+d_{R}+j}^{\dagger}c_{m_{0}+d_{R}+m}\right\rangle  & \left\langle c_{m_{0}+d_{R}+j}^{\dagger}c_{-m_{0}-d_{L}-m}\right\rangle \\
\left\langle c_{-m_{0}-d_{L}-j}^{\dagger}c_{m_{0}+d_{R}+m}\right\rangle  & \left\langle c_{-m_{0}-d_{L}-j}^{\dagger}c_{-m_{0}-d_{L}-m}\right\rangle 
\end{array}\right),
\end{equation}
such that $\left(C_{A}\right)_{2j-2+\sigma_{1},2m-2+\sigma_{2}}=\left(\Phi_{jm}\right)_{\sigma_{1}\sigma_{2}}$
for $\sigma_{1},\sigma_{2}\in\left\{ 1,2\right\} $. Substituting
the long-range approximation from Eq.~(\ref{eq:Correlation-function-large-distance}),
we have a block-Toeplitz structure of $C_{A}$, meaning that the entries
of the blocks $\Phi_{jm}$ depend on $j,m$ only through $j-m$. In
particular, we may define a $2\times2$ block-symbol $\Phi\!\left(k\right)$
such that
\begin{equation}
\Phi_{jm}\approx\int_{-\pi}^{\pi}\frac{dk}{2\pi}\Phi\!\left(k\right)e^{-i\left(j-m\right)k}.
\end{equation}
The block-symbol $\Phi\!\left(k\right)$ has four discontinuity points
at $k=\pm k_{{\scriptscriptstyle F,L}}$ and $k=\pm k_{{\scriptscriptstyle F,R}}$,
and has a different form depending on whether $k_{{\scriptscriptstyle F,L}}>k_{{\scriptscriptstyle F,R}}$
or $k_{{\scriptscriptstyle F,L}}<k_{{\scriptscriptstyle F,R}}$; we
analyze the former case for concreteness, and subsequently mention
how the results of the different steps are changed in the latter case.
For $k_{{\scriptscriptstyle F,L}}>k_{{\scriptscriptstyle F,R}}$,
the entries of $\Phi\!\left(k\right)$ are given by
\begin{align}
\Phi_{11}\!\left(k\right) & =\begin{cases}
1 & -k_{{\scriptscriptstyle F,R}}<k<k_{{\scriptscriptstyle F,R}},\\
{\cal T}\!\left(k\right) & k_{{\scriptscriptstyle F,R}}<k<k_{{\scriptscriptstyle F,L}},\\
0 & \text{otherwise},
\end{cases}\nonumber \\
\Phi_{22}\!\left(k\right) & =\begin{cases}
1 & -k_{{\scriptscriptstyle F,L}}<k<k_{{\scriptscriptstyle F,R}},\\
{\cal R}\!\left(k\right) & k_{{\scriptscriptstyle F,R}}<k<k_{{\scriptscriptstyle F,L}},\\
0 & \text{otherwise},
\end{cases}\nonumber \\
\Phi_{12}\!\left(k\right) & =\begin{cases}
t_{{\scriptscriptstyle L}}^{*}\!\left(k\right)r_{{\scriptscriptstyle L}}\!\left(k\right)e^{i\left(d_{L}-d_{R}\right)k} & k_{{\scriptscriptstyle F,R}}<k<k_{{\scriptscriptstyle F,L}},\\
0 & \text{otherwise},
\end{cases}\label{eq:Correlation-block-Toeplitz-symbol}
\end{align}
and $\Phi_{21}\!\left(k\right)=\Phi_{12}\!\left(k\right)^{*}$.

Following the customary method~\citep{Jin2004}, we write R\'enyi
entropies in terms of contour integrals, using the notation $D_{\ell}\!\left(\lambda\right)=\det\!\left(\lambda\mathbb{I}_{2\ell}-C_{A}\right)$:
\begin{equation}
S_{A}^{\left(n\right)}=\frac{1}{\left(1-n\right)2\pi i}\lim_{\varepsilon,\delta\to0^{+}}\int_{c\left(\varepsilon,\delta\right)}d\lambda\ln\left[\left(\lambda+\varepsilon\right)^{n}+\left(1+\varepsilon-\lambda\right)^{n}\right]\frac{d}{d\lambda}\ln D_{\ell}\!\left(\lambda\right).\label{eq:Renyi-contour-integral}
\end{equation}
Here $c\left(\varepsilon,\delta\right)$ is a closed contour that
encircles the segment $\left[0,1\right]$ (which contains all the
zeros of $D_{\ell}\!\left(\lambda\right)$), converges to this segment
as $\varepsilon,\delta\to0^{+}$, and avoids the singularities of
$\ln\!\left[\left(\lambda+\varepsilon\right)^{n}+\left(1+\varepsilon-\lambda\right)^{n}\right]$
(cf.~Ref.~\citep{10.21468/SciPostPhys.11.4.085}). Observing that
$\lambda\mathbb{I}_{2\ell}-C_{A}$ is also a block-Toeplitz matrix
(with respect to the block-symbol $\lambda\mathbb{I}_{2}-\Phi\!\left(k\right)$),
we use the result of the generalized Fisher-Hartwig conjecture for
the large-$\ell$ asymptotics of $\ln D_{\ell}\!\left(\lambda\right)$~\citep{PhysRevA.92.042334,PhysRevA.97.062301}:
\begin{equation}
\ln D_{\ell}\!\left(\lambda\right)\sim\ell\int_{-\pi}^{\pi}\frac{dk}{2\pi}\ln\det\!\left(\lambda\mathbb{I}_{2}-\Phi\!\left(k\right)\right)+\frac{\ln\ell}{4\pi^{2}}\sum_{r}{\rm Tr}\left[\ln^{2}\!\left[\left(\lambda\mathbb{I}_{2}-\Phi\!\left(k_{r}^{-}\right)\right)\left(\lambda\mathbb{I}_{2}-\Phi\!\left(k_{r}^{+}\right)\right)^{-1}\right]\right]+\ldots,
\end{equation}
where $k_{r}$ are the discontinuity points of $\Phi\!\left(k\right)$,
and where the ellipsis stands for a subleading constant term, along
with additional terms that vanish for $\ell\to\infty$. This then
yields
\begin{align}
\ln D_{\ell}\!\left(\lambda\right) & \sim\left[\frac{k_{{\scriptscriptstyle F,R}}}{\pi}\ln\left(\lambda-1\right)^{2}+\frac{k_{{\scriptscriptstyle F,L}}-k_{{\scriptscriptstyle F,R}}}{\pi}\ln\lambda\left(\lambda-1\right)+\frac{\pi-k_{{\scriptscriptstyle F,L}}}{\pi}\ln\lambda^{2}\right]\ell+\frac{1}{\pi^{2}}\left(\ln\frac{\lambda-1}{\lambda}\right)^{2}\ln\ell.\label{eq:Block-Toeplitz-determinant-asymptotics}
\end{align}
For $k_{{\scriptscriptstyle F,L}}<k_{{\scriptscriptstyle F,R}}$ we
get the same asymptotics of Eq.~(\ref{eq:Block-Toeplitz-determinant-asymptotics})
up to the replacement $L\leftrightarrow R$.

Crucially, the use of the Fisher-Hartwig asymptotic formula implicitly
assumes that $\ell$ is the largest length scale other than those
already taken to infinity. In particular, Eq.~(\ref{eq:Block-Toeplitz-determinant-asymptotics})
is valid only in the regime where $\ell\gg\left|d_{{\scriptscriptstyle L}}-d_{{\scriptscriptstyle R}}\right|$;
in other words, it applies solely to a symmetric configuration of
the intervals, such that they have both equal length and (approximately)
equal distance from the impurity. We can also examine the opposite
case $\ell\ll\left|d_{{\scriptscriptstyle L}}-d_{{\scriptscriptstyle R}}\right|$
by taking this limit in the block-Toeplitz matrix before invoking
the asymptotic formula for its determinant; this simply amounts to
setting $\Phi_{12}\left(k\right)=0$ for all $k$ in Eq.~(\ref{eq:Correlation-block-Toeplitz-symbol}).
For $k_{{\scriptscriptstyle F,L}}>k_{{\scriptscriptstyle F,R}}$,
the asymptotics of the block-Toeplitz determinant is then
\begin{align}
\ln D_{\ell}\!\left(\lambda\right) & \sim\ell\left[\frac{k_{{\scriptscriptstyle F,R}}}{\pi}\ln\left(\lambda-1\right)^{2}+\frac{k_{{\scriptscriptstyle F,L}}-k_{{\scriptscriptstyle F,R}}}{2\pi}\ln\lambda\left(\lambda-1\right)+\int_{k_{F,R}}^{k_{F,L}}\frac{dk}{2\pi}\ln\!\left[\left(\lambda-{\cal T}\!\left(k\right)\right)\left(\lambda-{\cal R}\!\left(k\right)\right)\right]+\frac{\pi-k_{{\scriptscriptstyle F,L}}}{\pi}\ln\lambda^{2}\right]\nonumber \\
 & +\frac{\ln\ell}{2\pi^{2}}\left(\ln\frac{\lambda-1}{\lambda}\right)^{2}\nonumber \\
 & +\frac{\ln\ell}{4\pi^{2}}\left[\left(\ln\frac{\lambda}{\lambda-{\cal T}\!\left(k_{{\scriptscriptstyle F,L}}\right)}\right)^{2}+\left(\ln\frac{\lambda}{\lambda-{\cal R}\!\left(k_{{\scriptscriptstyle F,L}}\right)}\right)^{2}+\left(\ln\frac{\lambda-1}{\lambda-{\cal T}\!\left(k_{{\scriptscriptstyle F,R}}\right)}\right)^{2}+\left(\ln\frac{\lambda-1}{\lambda-{\cal R}\!\left(k_{{\scriptscriptstyle F,R}}\right)}\right)^{2}\right],\label{eq:Block-Toeplitz-determinant-NoMirrorOverlap}
\end{align}
where again the result for $k_{{\scriptscriptstyle F,L}}<k_{{\scriptscriptstyle F,R}}$
is the same up to the exchange $L\leftrightarrow R$.

Substituting Eq.~(\ref{eq:Block-Toeplitz-determinant-asymptotics})
into Eq.~(\ref{eq:Renyi-contour-integral}), we observe that for
$\ell\gg\left|d_{{\scriptscriptstyle L}}-d_{{\scriptscriptstyle R}}\right|$
the linear term of $S_{A}^{\left(n\right)}$ vanishes, so that the
leading asymptotics is given by
\begin{align}
S_{A}^{\left(n\right)} & \sim\frac{4}{1-n}Q_{n}\!\left(0\right)\ln\ell=\frac{1+n}{3n}\ln\ell.\label{eq:Renyi-entropy-asymptotics-symmetric}
\end{align}
Interestingly, this is precisely the value $S_{A}^{\left(n\right)}$
would acquire in the absence of a scatterer, i.e., in a homogeneous
system of free fermions \citep{Calabrese_disjoint_2009}. In this
symmetric configuration, the effect of the scatterer on the entanglement
of $A$ with its complement is thus completely eliminated. This is
suggestive of the existence of a certain procedure -- whereby a ``folding''
transformation about the location of the impurity joins together mirroring
sites \citep{10.21468/SciPostPhys.14.4.070} -- that could be used
more generally to simplify the entanglement structure of similar steady
states. In contrast, for $\ell\ll\left|d_{{\scriptscriptstyle L}}-d_{{\scriptscriptstyle R}}\right|$
the R\'enyi entropies are given by
\begin{align}
S_{A}^{\left(n\right)} & \sim\frac{\ell}{1-n}\int_{k_{-}}^{k_{+}}\frac{dk}{\pi}\ln\left[\left({\cal T}\!\left(k\right)\right)^{n}+\left({\cal R}\!\left(k\right)\right)^{n}\right]+\frac{1+n}{6n}\ln\ell\nonumber \\
 & +\frac{\ln\ell}{1-n}\left[Q_{n}\!\left({\cal T}\!\left(k_{{\scriptscriptstyle F,L}}\right)\right)+Q_{n}\!\left({\cal R}\!\left(k_{{\scriptscriptstyle F,L}}\right)\right)+Q_{n}\!\left({\cal T}\!\left(k_{{\scriptscriptstyle F,R}}\right)\right)+Q_{n}\!\left({\cal R}\!\left(k_{{\scriptscriptstyle F,R}}\right)\right)\right],
\end{align}
as obtained from substituting Eq.~(\ref{eq:Block-Toeplitz-determinant-NoMirrorOverlap})
into Eq.~(\ref{eq:Renyi-contour-integral}).

We conclude that for $\ell_{{\scriptscriptstyle L}}=\ell_{{\scriptscriptstyle R}}\equiv\ell$,
and up to a logarithmic order, 
\begin{equation}
S_{A}^{\left(n\right)}=\begin{cases}
\frac{1+n}{3n}\ln\ell & \ell\gg\left|d_{{\scriptscriptstyle L}}-d_{{\scriptscriptstyle R}}\right|,\\
S_{A_{L}}^{\left(n\right)}+S_{A_{R}}^{\left(n\right)} & \ell\ll\left|d_{{\scriptscriptstyle L}}-d_{{\scriptscriptstyle R}}\right|,
\end{cases}
\end{equation}
where we used the single-interval results from Eq.~(\ref{eq:Renyi-entropies-single-intervals}).
Thus, for $\ell\gg\left|d_{{\scriptscriptstyle L}}-d_{{\scriptscriptstyle R}}\right|$
we find that the R\'enyi MI scales as
\begin{align}
{\cal I}_{A_{L}:A_{R}}^{\left(n\right)} & \sim\frac{\ell}{1-n}\int_{k_{-}}^{k_{+}}\frac{dk}{\pi}\ln\left[\left({\cal T}\!\left(k\right)\right)^{n}+\left({\cal R}\!\left(k\right)\right)^{n}\right]-\frac{1+n}{6n}\ln\ell\nonumber \\
 & +\frac{\ln\ell}{1-n}\left[Q_{n}\!\left({\cal T}\!\left(k_{{\scriptscriptstyle F,L}}\right)\right)+Q_{n}\!\left({\cal R}\!\left(k_{{\scriptscriptstyle F,L}}\right)\right)+Q_{n}\!\left({\cal T}\!\left(k_{{\scriptscriptstyle F,R}}\right)\right)+Q_{n}\!\left({\cal R}\!\left(k_{{\scriptscriptstyle F,R}}\right)\right)\right],\label{eq:Renyi-MI-symmetric}
\end{align}
while for $\ell\ll\left|d_{{\scriptscriptstyle L}}-d_{{\scriptscriptstyle R}}\right|$
the R\'enyi MI obeys an area-law scaling with $\ell$.

In Subsec.~\ref{subsec:Main-results} we emphasized that our analysis
relies on the assumption that $k_{{\scriptscriptstyle F,L}}\neq k_{{\scriptscriptstyle F,R}}$,
meaning that the analytical expressions for the logarithmic terms
of correlation measures cannot be naively taken to the no-bias limit.
This is true in particular for Eq.~(\ref{eq:Renyi-MI-symmetric}),
and it is instructive to explain why a simple substitution of $k_{{\scriptscriptstyle F,L}}=k_{{\scriptscriptstyle F,R}}$
fails in this case (i.e., why the limits do not commute). Indeed,
when $k_{{\scriptscriptstyle F,L}}=k_{{\scriptscriptstyle F,R}}$
the analysis of the Fisher-Hartwig asymptotics of $\ln D_{\ell}\!\left(\lambda\right)$
must be modified, since instead of the four discontinuity points that
the block-Toeplitz symbol in Eq.~(\ref{eq:Correlation-block-Toeplitz-symbol})
has in the presence of a bias, in the absence of a bias it has only
two discontinuity points. Furthermore, the symbol becomes independent
of the scattering matrix and of $d_{{\scriptscriptstyle L}}-d_{{\scriptscriptstyle R}}$.
It is straightforward to check that the asymptotics of the determinant
$\ln D_{\ell}\!\left(\lambda\right)$ is then equal to that appearing
in Eq.~(\ref{eq:Block-Toeplitz-determinant-asymptotics}) when $k_{{\scriptscriptstyle F,L}}=k_{{\scriptscriptstyle F,R}}$
is substituted into the equation. Therefore, Eq.~(\ref{eq:Renyi-entropy-asymptotics-symmetric})
captures the R\'enyi entropy of $A$ (regardless of the value of
$\left|d_{{\scriptscriptstyle L}}-d_{{\scriptscriptstyle R}}\right|$).
Combined with the fact that the entropy of each interval is given
by $S_{A_{i}}^{\left(n\right)}\sim\frac{1+n}{6n}\ln\ell$ (as was
already mentioned below Eq.~(\ref{eq:Renyi-entropies-single-intervals})),
we conclude that for $k_{{\scriptscriptstyle F,L}}=k_{{\scriptscriptstyle F,R}}$
both the linear term and the logarithmic term of the R\'enyi MI vanish;
this agrees with our observation from Subsec.~\ref{subsec:Correlation-matrix},
which stated that, in the absence of a bias, the R\'enyi MI must
vanish identically (in the long-range limit).

\subsubsection{Simplified steady state for the general case\label{subsec:Simplified-steady-state}}

We now turn to treat the more general configuration where $\ell_{{\scriptscriptstyle L}}$
and $\ell_{{\scriptscriptstyle R}}$ are not necessarily equal and
with an arbitrary value of $d_{{\scriptscriptstyle L}}-d_{{\scriptscriptstyle R}}$
(but still in the long-range limit $d_{i}/\ell_{i}\to\infty$). As
before, we compute R\'enyi entropies using their relation to the
restricted correlation matrix of the subsystem of interest, given
in Eq.~(\ref{eq:Entropy_from_correlations}). Through a series expansion,
this relation can be expressed as
\begin{equation}
S_{X}^{\left(n\right)}=\frac{1}{1-n}\sum_{s=1}^{\infty}\frac{\left(-1\right)^{s+1}}{s}{\rm Tr}\!\left[\left\{ \left(C_{X}\right)^{n}+\left(\mathbb{I}-C_{X}\right)^{n}-\mathbb{I}\right\} ^{s}\right].\label{eq:Renyi-entropy-series-expansion}
\end{equation}
Thus, it suffices to find a systematic way to compute correlation
matrix moments of the form ${\rm Tr}\!\left[\left(C_{X}\right)^{p}\right]$
(for any integer $p$), provided that the result allows to then sum
up the series expansion of Eq.~(\ref{eq:Renyi-entropy-series-expansion}).
Indeed, this was the basis for the analytical calculation we reported
in Ref.~\citep{fraenkel2022extensive} for the leading-order asymptotics.
Yet in order to produce the logarithmic corrections, we must replace
the steady state we described with a simplified version of it, from
which we can nevertheless read off the entanglement structure of the
original steady state. 

The construction of the artificial simplified steady state is done
through two assumptions. We first assume that particles emerge only
from the reservoir with the higher chemical potential and occupy only
the single-particle eigenstates associated with the momenta $k_{-}\le\left|k\right|\le k_{+}$
(i.e., the states with $\left|k\right|<k_{-}$ are taken to be empty).
Our results for the R\'enyi MI in Eqs.~(\ref{eq:Renyi_MI_volume_law})
and (\ref{eq:Renyi-MI-symmetric}) indeed lead to the realization
that, up to the linear and logarithmic order, the correlations between
$A_{{\scriptscriptstyle L}}$ and $A_{{\scriptscriptstyle R}}$ are
generated solely by particles occupying the eigenstates corresponding
to $k_{-}\le\left|k\right|\le k_{+}$, motivating the assumption that
this artificial state has the same entanglement structure as the true
state. It should be noted that the values of $S_{A_{i}}^{\left(n\right)}$
to which this artificial steady state gives rise disagree at the logarithmic
order with the same quantities as computed for the true steady state
(as they appear in Eq.~(\ref{eq:Renyi-entropies-single-intervals}));
yet the result for $S_{A_{L}}^{\left(n\right)}+S_{A_{R}}^{\left(n\right)}$
is the same for the two states, and this is indeed the relevant quantity
for the MI calculation. The second assumption that we impose is that
the scattering amplitudes in Eq.~(\ref{eq:Sacttering-matrix}) are
$k$-independent; this will be amended later by an appropriate insertion
of the $k$-dependence into the final result that we derive.

\subsubsection{Expressing entropies with Toeplitz determinants\label{subsec:MI-Entropies-with-Toeplitz}}

We continue by leveraging the simplicity of the artificial steady
state to obtain useful formulae for the R\'enyi entropies of $A_{{\scriptscriptstyle L}}$,
$A_{{\scriptscriptstyle R}}$ and $A$, which will later produce the
R\'enyi MI asymptotics. Even though in the case of $A_{{\scriptscriptstyle L}}$
and $A_{{\scriptscriptstyle R}}$ the asymptotics of the entropies
are already known from Ref.~\citep{10.21468/SciPostPhys.11.4.085},
we include them in the analysis as well. This is because it will turn
out that treating directly the R\'enyi MI rather than the different
entropies comprising it is somewhat simpler, due to terms that are
canceled out when these entropies are combined. For concreteness,
we present the following analysis focusing on the case $k_{{\scriptscriptstyle F,L}}>k_{{\scriptscriptstyle F,R}}$.

In Appendix \ref{sec:Stationary-phase-approximation} we explain how
the simplified definition in Subsec.~\ref{subsec:Simplified-steady-state}
of the steady state leads, via a stationary phase approximation \citep{doi:10.1137/1.9780898719260},
to the following approximation of the correlation matrix moments:
\begin{equation}
{\rm Tr}\!\left[\left(C_{X}\right)^{p}\right]\approx\int_{\left[k_{F,R},k_{F,L}\right]^{p}}\frac{d^{p}k}{\left(2\pi\right)^{p}}\prod_{j=1}^{p}\left\{ \sum_{m\in X_{L}}\left[e^{im\left(k_{j-1}-k_{j}\right)}+{\cal R}e^{im\left(k_{j}-k_{j-1}\right)}\right]+\sum_{m\in X_{R}}{\cal T}e^{im\left(k_{j-1}-k_{j}\right)}\right\} .\label{eq:Approximated-correlation-moment}
\end{equation}
Here we define $k_{0}=k_{p}$, and we see each subsystem of interest
$X$ as comprised of two disjoint components -- $X_{{\scriptscriptstyle L}}$
to the left of the impurity and $X_{{\scriptscriptstyle R}}$ to its
right -- each being contiguous; if $X=A_{{\scriptscriptstyle L}}$
or $X=A_{{\scriptscriptstyle R}}$ then one of these components is
trivial. If we discretize the integral in Eq.~(\ref{eq:Approximated-correlation-moment})
by dividing the interval $\left[k_{{\scriptscriptstyle F,R}},k_{{\scriptscriptstyle F,L}}\right]$
into $M\gg1$ equal-length smaller intervals, we obtain
\begin{equation}
{\rm Tr}\!\left[\left(C_{X}\right)^{p}\right]\approx\left(\frac{\Delta k}{2\pi M}\right)^{p}\sum_{s_{1},\ldots,s_{p}=1}^{M}\prod_{j=1}^{p}\left\{ \sum_{m\in X_{L}}\left[e^{im\left(k_{s_{j-1}}-k_{s_{j}}\right)}+{\cal R}e^{im\left(k_{s_{j}}-k_{s_{j-1}}\right)}\right]+\sum_{m\in X_{R}}{\cal T}e^{im\left(k_{s_{j-1}}-k_{s_{j}}\right)}\right\} ,\label{eq:Riemann-sum-approximated-moment}
\end{equation}
where $\Delta k=k_{{\scriptscriptstyle F,L}}-k_{{\scriptscriptstyle F,R}}$
and $k_{s}=k_{{\scriptscriptstyle F,R}}+\frac{\Delta k}{M}s$. This
is merely an approximation of the integral as a Riemann sum, yet it
constitutes a crucial step on the way to our desired solution. We
observe that Eq.~(\ref{eq:Riemann-sum-approximated-moment}) is equivalent
to writing
\begin{equation}
{\rm Tr}\left[\left(C_{X}\right)^{p}\right]\approx\left(\frac{\Delta k}{2\pi M}\right)^{p}{\rm Tr}\left[\left(\widehat{K}^{\left(X\right)}\right)^{p}\right],
\end{equation}
with $\widehat{K}^{\left(X\right)}$ being a $M\times M$ Toeplitz
matrix, defined by
\begin{equation}
\left(\widehat{K}^{\left(X\right)}\right)_{ss'}=\sum_{m\in X_{L}}\left[e^{-im\Delta k\left(s-s'\right)/M}+{\cal R}e^{im\Delta k\left(s-s'\right)/M}\right]+\sum_{m\in X_{R}}{\cal T}e^{-im\Delta k\left(s-s'\right)/M}.\label{eq:Toeplitz-matrix-entries}
\end{equation}
Then, by plugging this approximation into Eq.~(\ref{eq:Renyi-entropy-series-expansion})
and summing the series, we find that
\begin{equation}
S_{X}^{\left(n\right)}\approx\frac{1}{1-n}{\rm Tr}\ln\!\left[\left(\frac{\Delta k}{2\pi M}\widehat{K}^{\left(X\right)}\right)^{n}+\left(\mathbb{I}-\frac{\Delta k}{2\pi M}\widehat{K}^{\left(X\right)}\right)^{n}\right].
\end{equation}

To complete this part, we introduce the following notation for any
integer $n\ge2$: 
\begin{equation}
\left(z_{\gamma}\right)^{-1}=1-e^{2\pi i\gamma/n},\,\,\,\,\,\gamma=-\frac{n-1}{2},-\frac{n-3}{2},\ldots,\frac{n-1}{2}.
\end{equation}
Note that if $n$ is odd then one of the possible values of $\gamma$
is $\gamma=0$, so that $\left(z_{\gamma=0}\right)^{-1}=0$. The points
$z_{\gamma}$ are the roots of the polynomial
\begin{equation}
p_{n}\!\left(z\right)=z^{n}+\left(1-z\right)^{n}=\prod_{\gamma=-\frac{n-1}{2}}^{\frac{n-1}{2}}\left(1-\frac{z}{z_{\gamma}}\right).\label{eq:MI-polynomial-definition}
\end{equation}
For an even $n$, $p_{n}$ has $n$ different roots, while for an
odd $n$, $p_{n}$ is of degree $n-1$ and has only $n-1$ different
roots, with $\gamma=0$ designating the missing root. This decomposition
of the polynomial $p_{n}$ allows us to write
\begin{equation}
S_{X}^{\left(n\right)}\approx\frac{1}{1-n}\sum_{\gamma=-\frac{n-1}{2}}^{\frac{n-1}{2}}\ln{\cal Z}_{\gamma}^{\left(X\right)},\label{eq:Renyi-entropies-as-log_determinant-sums}
\end{equation}
where each ${\cal Z}_{\gamma}^{\left(X\right)}=\det K_{\gamma}^{\left(X\right)}$
is a Toeplitz determinant, with the corresponding Toeplitz matrix
being $\left(K_{\gamma}^{\left(X\right)}\right)_{ss'}=\delta_{ss'}+\frac{\Delta k}{2\pi M}\left(e^{2\pi i\gamma/n}-1\right)\left(\widehat{K}^{\left(X\right)}\right)_{ss'}$.

Eq.~(\ref{eq:Renyi-entropies-as-log_determinant-sums}) represents
a consequential step in our calculation, as it directly expresses
the R\'enyi entropies using (logarithms of) Toeplitz determinants,
the asymptotics of which are captured by the Fisher-Hartwig formula.
The large parameter $M$ determining the asymptotic regime is not
an intrinsic quantity related to the model or the state in question,
but rather a fictitious quantity arising from our discretization of
the momentum coordinates. We can therefore always choose it to be
arbitrarily large, and the limit $M\to\infty$ must be eventually
taken to recover the exact expressions for the entropies.

\subsubsection{Asymptotics of entropies via the Fisher-Hartwig formula\label{subsec:MI-Fisher-Hartwig-asymptotics}}

In order to make use of the Fisher-Hartwig formula for the large-$M$
asymptotics of the Toeplitz determinants appearing in Eq.~(\ref{eq:Renyi-entropies-as-log_determinant-sums}),
we must first cast the entries of the Toeplitz matrices $K_{\gamma}^{\left(X\right)}$
in the following form:
\begin{equation}
\left(K_{\gamma}^{\left(X\right)}\right)_{ss'}=\int_{-\pi}^{\pi}\frac{d\theta}{2\pi}e^{-i\left(s-s'\right)\theta}\phi_{\gamma}^{\left(X\right)}\!\left(\theta\right),\label{eq:Toeplitz-matrix-expression-with-symbol}
\end{equation}
where the function $\phi_{\gamma}^{\left(X\right)}\!\left(\theta\right)$
is the appropriate Toeplitz symbol. For this purpose, we define for
$i=L,R$ the angles
\begin{equation}
\theta_{i,-}=\frac{d_{i}}{M}\Delta k,\,\,\,\,\,\theta_{i,+}=\frac{d_{i}+\ell_{i}}{M}\Delta k,\label{eq:Toeplitz-symbol-discontinuities}
\end{equation}
and choose $M$ such that $\theta_{i,\pm}$ are not negligible but
smaller than $\pi$ (the latter simply requires $M>d_{i}+\ell_{i}$).
For $\ell_{i}\gg1$, we can then approximate the the entries of $K_{\gamma}^{\left(X\right)}$
as integrals over $\left[-\pi,\pi\right]$, by replacing $m\Delta k/M$
in Eq.~(\ref{eq:Toeplitz-matrix-entries}) with a continuous integration
variable $\theta$, yielding
\begin{equation}
\phi_{\gamma}^{\left(X\right)}\!\left(\theta\right)\approx\begin{cases}
T_{\gamma}^{\left(X\right)}\!\left(\theta\right) & -\pi\le\theta<0,\\
{\cal T}\cdot T_{\gamma}^{\left(X\right)}\!\left(\theta\right)+{\cal R}\cdot T_{\gamma}^{\left(X\right)}\!\left(-\theta\right) & 0\le\theta<\pi,
\end{cases}\label{eq:Toeplitz-symbol}
\end{equation}
where
\begin{align}
T_{\gamma}^{\left(A_{L}\right)}\!\left(\theta\right) & =\begin{cases}
e^{2\pi i\gamma/n} & \theta\in\left[-\theta_{{\scriptscriptstyle L},+},-\theta_{{\scriptscriptstyle L},-}\right],\\
1 & {\rm otherwise}\,,
\end{cases}\nonumber \\
T_{\gamma}^{\left(A_{R}\right)}\!\left(\theta\right) & =\begin{cases}
e^{2\pi i\gamma/n} & \theta\in\left[\theta_{{\scriptscriptstyle R},-},\theta_{{\scriptscriptstyle R},+}\right],\\
1 & {\rm otherwise}\,,
\end{cases}\nonumber \\
T_{\gamma}^{\left(A\right)}\!\left(\theta\right) & =\begin{cases}
e^{2\pi i\gamma/n} & \theta\in\left[-\theta_{{\scriptscriptstyle L},+},-\theta_{{\scriptscriptstyle L},-}\right]\cup\left[\theta_{{\scriptscriptstyle R},-},\theta_{{\scriptscriptstyle R},+}\right],\\
1 & {\rm otherwise}\,.
\end{cases}
\end{align}

The Toeplitz symbol $\phi_{\gamma}^{\left(X\right)}$ in Eq.~(\ref{eq:Toeplitz-symbol})
can be cast in the Fisher-Hartwig form~\citep{10.2307/23030524},
namely $\phi_{\gamma}^{\left(X\right)}\!\left(\theta\right)=\prod_{r}g_{r}\!\left(\theta\right)$,
where the index $r$ is associated with the discontinuity points $\theta_{r}$
of $\phi_{\gamma}^{\left(X\right)}$, and where 
\begin{align}
g_{r}\!\left(\theta\right) & =\begin{cases}
e^{i\pi\beta_{r}} & -\pi\le\theta<\theta_{r},\\
e^{-i\pi\beta_{r}} & \theta_{r}\le\theta<\pi,
\end{cases}
\end{align}
using the notation
\begin{equation}
\beta_{r}=\frac{1}{2\pi i}\ln\!\left(\frac{\phi_{\gamma}^{\left(X\right)}\!\left(\theta_{r}^{-}\right)}{\phi_{\gamma}^{\left(X\right)}\!\left(\theta_{r}^{+}\right)}\right).
\end{equation}
The branch of the logarithm is always chosen such that $\left|{\rm Im}\!\left[\ln z\right]\right|<\pi$,
implying that $\left|{\rm Re}\!\left[\beta_{r}\right]\right|<\frac{1}{2}$
and thus allowing the direct use of the Fisher-Hartwig asymptotic
formula~\citep{10.2307/23030524}. For $M\gg1$ this formula then
gives
\begin{equation}
\ln{\cal Z}_{\gamma}^{\left(X\right)}\sim\left[i\sum_{r}\beta_{r}\theta_{r}\right]M+\left[-\sum_{r}\beta_{r}^{2}\right]\ln M+2\sum_{r_{1}<r_{2}}\beta_{r_{1}}\beta_{r_{2}}\ln\!\left|e^{i\theta_{r_{2}}}-e^{i\theta_{r_{1}}}\right|+\ldots,\label{eq:Fisher-Hartwig-asymptotics-large-M}
\end{equation}
where another constant-in-$M$ term that is independent of $\theta_{r}$
(and thus uninformative regarding the scaling with $\ell_{i}$ and
$d_{i}$) was absorbed into the ellipsis, along with $o\!\left(1\right)$
terms. As the symbol $\phi_{\gamma}^{\left(X\right)}$ is piecewise-constant
and satisfies $\sum_{r}\beta_{r}=0$, Eq.~(\ref{eq:Fisher-Hartwig-asymptotics-large-M})
can be rewritten as
\begin{align}
\ln{\cal Z}_{\gamma}^{\left(X\right)} & \sim\frac{1}{2\pi}\left[\sum_{r}\left(\theta_{r+1}-\theta_{r}\right)\ln\!\left(\phi_{\gamma}^{\left(X\right)}\!\left(\theta_{r}^{+}\right)\right)\right]M-\frac{1}{2\pi^{2}}\sum_{r_{1}<r_{2}}\ln\!\left(\frac{\phi_{\gamma}^{\left(X\right)}\!\left(\theta_{r_{1}}^{-}\right)}{\phi_{\gamma}^{\left(X\right)}\!\left(\theta_{r_{1}}^{+}\right)}\right)\ln\!\left(\frac{\phi_{\gamma}^{\left(X\right)}\!\left(\theta_{r_{2}}^{-}\right)}{\phi_{\gamma}^{\left(X\right)}\!\left(\theta_{r_{2}}^{+}\right)}\right)\ln\!\left|M\!\left(e^{i\theta_{r_{2}}}-e^{i\theta_{r_{1}}}\right)\right|.\label{eq:Fisher-Hartwig-asymptotics-simplified}
\end{align}

We now examine the asymptotics of the R\'enyi MI that results from
Eq.~(\ref{eq:Fisher-Hartwig-asymptotics-simplified}). For this purpose
we define
\begin{equation}
\ln{\cal Z}_{\gamma}^{\left({\cal I}\right)}=\ln{\cal Z}_{\gamma}^{\left(A_{L}\right)}+\ln{\cal Z}_{\gamma}^{\left(A_{R}\right)}-\ln{\cal Z}_{\gamma}^{\left(A\right)},\label{eq:MI-gamma-components}
\end{equation}
such that, through Eq.~(\ref{eq:Renyi-entropies-as-log_determinant-sums}),
the R\'enyi MI is given by ${\cal I}_{A_{L}:A_{R}}^{\left(n\right)}\approx\frac{1}{1-n}\sum_{\gamma}\ln{\cal Z}_{\gamma}^{\left({\cal I}\right)}$.
We note that the symbol $\phi_{\gamma}^{\left(A_{L}\right)}$ has
4 discontinuities at $-\theta_{{\scriptscriptstyle L},+},-\theta_{{\scriptscriptstyle L},-},\theta_{{\scriptscriptstyle L},-},\theta_{{\scriptscriptstyle L},+}$;
the symbol $\phi_{\gamma}^{\left(A_{R}\right)}$ has 2 discontinuities
at $\theta_{{\scriptscriptstyle R},-},\theta_{{\scriptscriptstyle R},+}$;
and the symbol $\phi_{\gamma}^{\left(A\right)}$ has, in principle,
6 discontinuities at all of these points, unless some coincide (but
it has at least 4 discontinuities). In the derivation that follows,
we will assume that $\phi_{\gamma}^{\left(A\right)}$ has 6 different
discontinuities, and address the degenerate cases after completing
our argument. We reiterate that we will eventually take the limits
$M\to\infty$ (replacing Riemann sums with integrals) and $d_{i}\to\infty$
(the long-range limit).

For convenience, we separate $\ln{\cal Z}_{\gamma}^{\left({\cal I}\right)}$
into two terms,
\begin{equation}
\ln{\cal Z}_{\gamma}^{\left({\cal I}\right)}\sim{\cal C}_{\gamma}^{\left({\rm lin}\right)}\ell_{{\rm mirror}}+{\cal G}_{\gamma}^{\left({\rm log}\right)}\!\left(\ell_{{\scriptscriptstyle L}},\ell_{{\scriptscriptstyle R}},d_{{\scriptscriptstyle L}}-d_{{\scriptscriptstyle R}}\right),\label{eq:MI-scaling-components}
\end{equation}
where ${\cal C}_{\gamma}^{\left({\rm lin}\right)}$ is constant in
the different length scales (such that ${\cal C}_{\gamma}^{\left({\rm lin}\right)}\ell_{{\rm mirror}}$
is the volume-law term of $\ln{\cal Z}_{\gamma}^{\left({\cal I}\right)}$),
while ${\cal G}_{\gamma}^{\left({\rm log}\right)}$ is logarithmic
in combinations of $\ell_{{\scriptscriptstyle L}},\ell_{{\scriptscriptstyle R}},d_{{\scriptscriptstyle L}}-d_{{\scriptscriptstyle R}}$.
When combining the contributions from the different subsystems, the
term in Eq.~(\ref{eq:Fisher-Hartwig-asymptotics-simplified}) that
is linear in $M$ produces the linear-in-$\ell_{{\rm mirror}}$ term
in Eq.~(\ref{eq:MI-scaling-components}), while the second term in
Eq.~(\ref{eq:Fisher-Hartwig-asymptotics-simplified}) yields ${\cal G}_{\gamma}^{\left({\rm log}\right)}$.

A straightforward calculation shows that ${\cal C}_{\gamma}^{\left({\rm lin}\right)}$
is independent of $M$ and $d_{i}$ even before the limits $M\to\infty$
and $d_{{\scriptscriptstyle i}}\to\infty$ are taken; namely, we obtain
\begin{equation}
{\cal C}_{\gamma}^{\left({\rm lin}\right)}=\Delta k\left[\frac{1}{2\pi}\ln\!\left(1-\frac{{\cal T}}{z_{\gamma}}\right)+\frac{1}{2\pi}\ln\!\left(1-\frac{{\cal {\cal R}}}{z_{\gamma}}\right)-\frac{i\gamma}{n}\right].
\end{equation}
Using Eq.~(\ref{eq:MI-polynomial-definition}), the summation over
$\gamma$ then gives
\begin{equation}
\sum_{\gamma=-\frac{n-1}{2}}^{\frac{n-1}{2}}{\cal C}_{\gamma}^{\left({\rm lin}\right)}=\frac{\Delta k}{\pi}\ln\!\left({\cal T}^{n}+{\cal R}^{n}\right).\label{eq:MI-linear-term-k-independent}
\end{equation}
It is easy to check that if we define $\Delta k=\left|k_{{\scriptscriptstyle F,L}}-k_{{\scriptscriptstyle F,R}}\right|$,
then the result in Eq.~(\ref{eq:MI-linear-term-k-independent}) applies
also in the case $k_{{\scriptscriptstyle F,L}}<k_{{\scriptscriptstyle F,R}}$.
As a consistency check, we observe that Eq.~(\ref{eq:MI-linear-term-k-independent})
precisely matches what appears in Eq.~(\ref{eq:Renyi_MI_volume_law})
if one assumes $k$-independent scattering probabilities.

As for the logarithmic contribution ${\cal G}_{\gamma}^{\left({\rm log}\right)}$
in Eq.~(\ref{eq:MI-scaling-components}), it arises from a sum over
``interactions'' between discontinuities of the Toeplitz symbols,
as evident from Eq.~(\ref{eq:Fisher-Hartwig-asymptotics-simplified}).
These ``interactions'' vary with the position of $\bar{A}_{{\scriptscriptstyle L}}=\left\{ m|-m\in A_{{\scriptscriptstyle L}}\right\} $,
the mirror image of $A_{{\scriptscriptstyle L}}$, relative to $A_{{\scriptscriptstyle R}}$.
By recalling that the limit $M\to\infty$ is the next to be taken,
and given that the angles $\theta_{r}$ all tend to zero in this limit
(see Eq.~(\ref{eq:Toeplitz-symbol-discontinuities})), we can justifiably
replace $M\!\left(e^{i\theta_{r_{2}}}-e^{i\theta_{r_{1}}}\right)$
in Eq.~(\ref{eq:Fisher-Hartwig-asymptotics-simplified}) with $M\!\left(\theta_{r_{2}}-\theta_{r_{1}}\right)$,
which is independent of $M$.

We additionally observe that the discontinuities at $\theta=-\theta_{{\scriptscriptstyle L},\pm}$
of $\phi_{\gamma}^{\left(A_{L}\right)}$ and $\phi_{\gamma}^{\left(A\right)}$
may be disregarded in this computation, due to the following reason.
The ``interaction'' between these two points (i.e., the summand
with $\theta_{r_{1}}=-\theta_{{\scriptscriptstyle L},+}$ and $\theta_{r_{2}}=-\theta_{{\scriptscriptstyle L},-}$
in Eq.~(\ref{eq:Fisher-Hartwig-asymptotics-simplified})) is canceled
out in the MI due to an identical contribution from $\ln{\cal Z}_{\gamma}^{\left(A_{L}\right)}$
and $\ln{\cal Z}_{\gamma}^{\left(A\right)}$ in Eq.~(\ref{eq:MI-gamma-components}).
Moreover, we observe that for $X=A_{{\scriptscriptstyle L}},A$,
\begin{equation}
\ln\!\left[\frac{\phi_{\gamma}^{\left(X\right)}\!\left(\left(-\theta_{{\scriptscriptstyle L},+}\right)^{-}\right)}{\phi_{\gamma}^{\left(X\right)}\!\left(\left(-\theta_{{\scriptscriptstyle L},+}\right)^{+}\right)}\right]=-\frac{2\pi i\gamma}{n}=-\ln\!\left[\frac{\phi_{\gamma}^{\left(X\right)}\!\left(\left(-\theta_{{\scriptscriptstyle L},-}\right)^{-}\right)}{\phi_{\gamma}^{\left(X\right)}\!\left(\left(-\theta_{{\scriptscriptstyle L},-}\right)^{+}\right)}\right],
\end{equation}
so that the ``interaction'' of any discontinuity $\theta_{r}>0$
with these two points will contribute in total to ${\cal G}_{\gamma}^{\left({\rm log}\right)}$
a term of the form $\frac{i\gamma}{\pi n}\ln\!\left(\phi_{\gamma}\!\left(\theta_{r}^{-}\right)/\phi_{\gamma}\!\left(\theta_{r}^{+}\right)\right)\ln\!\left|\frac{\theta_{r}+\theta_{{\scriptscriptstyle L},+}}{\theta_{r}+\theta_{{\scriptscriptstyle L},-}}\right|$.
Since $\left|\frac{\theta_{r}+\theta_{{\scriptscriptstyle L},+}}{\theta_{r}+\theta_{{\scriptscriptstyle L},-}}\right|\to1$
in the limit $d_{i}\to\infty$ (which will be eventually taken), this
contribution will always vanish in the long-range limit.

What is therefore left to do is to sum over the contributions of ``interactions''
between the four discontinuities at $\theta_{{\scriptscriptstyle L},\pm},\theta_{{\scriptscriptstyle R},\pm}$,
as well as to sum over the index $\gamma$. The result of this sum
turns out to have a different form depending on the location and size
of $\bar{A}_{{\scriptscriptstyle L}}$ relative to $A_{{\scriptscriptstyle R}}$.
That is, a separate calculation is required for each of the following
three cases: (i) one of the two intervals $\bar{A}_{{\scriptscriptstyle L}}$
and $A_{{\scriptscriptstyle R}}$ contains the other; (ii) the intervals
$\bar{A}_{{\scriptscriptstyle L}}$ and $A_{{\scriptscriptstyle R}}$
do not intersect; (iii) the intersection of $\bar{A}_{{\scriptscriptstyle L}}$
and $A_{{\scriptscriptstyle R}}$ is a proper subsystem of each of
them. The common rationale that these calculations all share is that
the discrete sum over $\gamma$, which includes $n$ terms, can be
expressed as a complex contour integral, by exploiting the residue
theorem with respect to the roots of the polynomial $p_{n}$ defined
in Eq.~(\ref{eq:MI-polynomial-definition}). This promotes $n$ from
an integer index to a parameter that can be varied continuously, which
is of course crucial to our ability to eventually take the limit $n\to1$
of the R\'enyi MI. We go over the details of these different calculations
in Appendix \ref{sec:Mutual-information-derivation-appendix}.

Fortunately, the results to which these separate calculations lead
can all be encapsulated in a single formula. Letting $m_{1}\le m_{2}\le m_{3}\le m_{4}$
denote the lengths $d_{{\scriptscriptstyle L}}$, $\ell_{{\scriptscriptstyle L}}+d_{{\scriptscriptstyle L}}$,
$d_{{\scriptscriptstyle R}}$ and $\ell_{{\scriptscriptstyle R}}+d_{{\scriptscriptstyle R}}$
in ascending order, we find that
\begin{align}
\sum_{\gamma=-\frac{n-1}{2}}^{\frac{n-1}{2}}{\cal G}_{\gamma}^{\left({\rm log}\right)} & =\left(Q_{n}\!\left({\cal T}\right)+Q_{n}\!\left({\cal R}\right)-\frac{1}{12}\left(\frac{1}{n}-n\right)\right)\ln\!\left|\frac{\left(m_{3}-m_{1}\right)\left(m_{4}-m_{2}\right)}{\left(\ell_{{\scriptscriptstyle L}}+d_{{\scriptscriptstyle L}}-\ell_{{\scriptscriptstyle R}}-d_{{\scriptscriptstyle R}}\right)\left(d_{{\scriptscriptstyle L}}-d_{{\scriptscriptstyle R}}\right)}\right|\nonumber \\
 & +\widetilde{Q}_{n}\!\left({\cal T}\right)\ln\!\left|\frac{\left(m_{3}-m_{1}\right)\left(m_{4}-m_{2}\right)}{\left(\ell_{{\scriptscriptstyle R}}+d_{{\scriptscriptstyle R}}-d_{{\scriptscriptstyle L}}\right)\left(\ell_{{\scriptscriptstyle L}}+d_{{\scriptscriptstyle L}}-d_{{\scriptscriptstyle R}}\right)}\right|,\label{eq:MI-log-term-k-independent}
\end{align}
where the functions $Q_{n}$ and $\widetilde{Q}_{n}$ were defined
in Eqs.~(\ref{eq:Log-scaling-kernel-overlap}) and (\ref{eq:Log-scaling-kernel-no-overlap}).

Before concluding this part, we address the fact that our derivation
relied on the symbols $\phi_{\gamma}^{\left(A\right)}$ having 6 discontinuity
points that do not coincide, i.e., on the assumption that $\left\{ \theta_{{\scriptscriptstyle L},\pm}\right\} \cap\left\{ \theta_{{\scriptscriptstyle R},\pm}\right\} =\phi$,
which corresponds (see Eq.~(\ref{eq:Toeplitz-symbol-discontinuities}))
to $\left\{ d_{{\scriptscriptstyle L}},d_{{\scriptscriptstyle L}}+\ell_{{\scriptscriptstyle L}}\right\} \cap\left\{ d_{{\scriptscriptstyle R}},d_{{\scriptscriptstyle R}}+\ell_{{\scriptscriptstyle R}}\right\} =\phi$.
Our argument readily extends to cases where such coincidences do in
fact occur. Indeed, examining the asymptotic formula in Eq.~(\ref{eq:Fisher-Hartwig-asymptotics-simplified})
for $X=A$, if two discontinuity points $\theta_{r_{1}}$ and $\theta_{r_{2}}$
coincide (i.e., $\theta_{r_{1}}=\theta_{r_{2}}$), then the same formula
can still be used if we only drop the divergent ``interaction''
term between $\theta_{r_{1}}$ and $\theta_{r_{2}}$, and then take
the limit $\theta_{r_{1}}\to\theta_{r_{2}}$; the set $\left\{ \theta_{r}\right\} $
keeps including all 6 points, even though two of them are now degenerate.
This logic entails that, in the asymptotics of $\ln{\cal Z}_{\gamma}^{\left(A\right)}$,
the degeneracy of $\theta_{r_{1}}$ and $\theta_{r_{2}}$ simply results
in omitting the term featuring $\ln\!\left|M\!\left(\theta_{r_{2}}-\theta_{r_{1}}\right)\right|$.
Finally, as such a degeneracy results from an equality between two
length scales out of the set $\left\{ d_{{\scriptscriptstyle L}},d_{{\scriptscriptstyle L}}+\ell_{{\scriptscriptstyle L}},d_{{\scriptscriptstyle R}},d_{{\scriptscriptstyle R}}+\ell_{{\scriptscriptstyle R}}\right\} $,
we infer that we may use Eq.~(\ref{eq:MI-log-term-k-independent})
even if such an equality holds, by simply dropping out of the logarithm
the problematic difference that formally vanishes. This justifies
the rule-of-thumb that was stated in this spirit in Subsec.~\ref{subsec:Main-results}.

\subsubsection{Going back to the true steady state}

As already emphasized, the linear and logarithmic contributions appearing
in Eqs.~(\ref{eq:MI-linear-term-k-independent}) and (\ref{eq:MI-log-term-k-independent})
were exactly computed for a simplified steady state, rather than for
the original steady state we were investigating. While the entanglement
structure should remain the same under the assumption that only the
single-particle states inside the voltage window are occupied, the
second assumption of $k$-independent scattering coefficients is indeed
more restrictive. The exact leading-order asymptotics of Eq.~(\ref{eq:Renyi_MI_volume_law})
shows how the volume-law term generalizes to the case of $k$-dependent
scattering. As for the subleading logarithmic contribution, we may
gain insight regarding a similar generalization by comparing Eq.~(\ref{eq:MI-log-term-k-independent})
to Eq.~(\ref{eq:Renyi-MI-symmetric}). We remind that the latter
was derived assuming a general $k$-dependent scattering matrix, but
was limited only for the symmetric case with $\ell_{{\scriptscriptstyle L}}=\ell_{{\scriptscriptstyle R}}$
and $d_{{\scriptscriptstyle L}}=d_{{\scriptscriptstyle R}}$. In this
symmetric case, an agreement between Eq.~(\ref{eq:MI-log-term-k-independent})
and Eq.~(\ref{eq:Renyi-MI-symmetric}) is achieved if we substitute
the two Fermi momenta $k_{{\scriptscriptstyle F,L}}$ and $k_{{\scriptscriptstyle F,R}}$
into the scattering probabilities appearing in Eq.~(\ref{eq:MI-log-term-k-independent}),
and then average over their contributions with equal weights.

The same extension of Eq.~(\ref{eq:MI-log-term-k-independent}) to
$k$-dependent scattering probabilities should be valid in general,
which is tantamount to the realization that the logarithmic contributions
to correlation measures arise from discontinuous jumps in the occupation
distribution of momentum states, and that the contributions from the
different jumps are weighted equally. This produces the logarithmic
term appearing in the final result reported in Eq.~(\ref{eq:Renyi-MI-full-asymptotics}).
This generalization is indeed confirmed numerically, as detailed in
Subsec.~\ref{subsec:Comparison-to-numerics}.

\subsection{Derivation of the negativity asymptotics\label{subsec:Derivation-of-negativity}}

The method that we devised for the calculation of the R\'enyi MI
asymptotics can be naturally generalized to facilitate the computation
of other quantities that are related to integer moments of two-point
correlation matrices. As already stated, a prominent example for such
a quantity is the R\'enyi negativity, which, for Gaussian states,
can indeed be expressed using the two-point correlation function (see
Eq.~(\ref{eq:Negativity_from_correlations})). Here we show how we
applied the method to extract the exact asymptotics of R\'enyi negativities
between $A_{{\scriptscriptstyle L}}$ and $A_{{\scriptscriptstyle R}}$,
leading us to Eq.~(\ref{eq:Renyi-negativity-full-asymptotics-symmetric}).
The description of the derivation is kept rather brief compared to
Subsec.~\ref{subsec:Derivation-of-MI}, given the close resemblance
between the steps of the two computations.

In Ref.~\citep{fraenkel2022extensive} we used the series expansion
of Eq.~(\ref{eq:Negativity_from_correlations_variation}) (with $X_{1}$
and $X_{2}$ standing for $A_{{\scriptscriptstyle L}}$ and $A_{{\scriptscriptstyle R}}$)
to write the $n$th R\'enyi negativity as
\begin{equation}
{\cal E}_{n}=\sum_{s=1}^{\infty}\frac{\left(-1\right)^{s+1}}{s}{\rm Tr}\!\left[\left\{ \prod_{\gamma=-\frac{n-1}{2}}^{\frac{n-1}{2}}\left(\mathbb{I}-C_{\gamma}\right)-\mathbb{I}\right\} ^{\!s}\,\right],
\end{equation}
and thus to show that its calculation may be reduced to that of the
joint moments ${\rm Tr}\!\left[C_{\gamma_{1}}C_{\gamma_{2}}\ldots C_{\gamma_{p}}\right]$,
for an arbitrary positive integer $p$. In this sense, the role of
the joint moments in the calculation of R\'enyi negativities is analogous
to the role of the correlation matrix moments in the calculation of
R\'enyi entropies (see Eq.~(\ref{eq:Renyi-entropy-series-expansion})).
Therefore, for the same simplified steady state that was defined in
Subsec.~\ref{subsec:Simplified-steady-state}, a procedure similar
to the one applied in Subsec.~\ref{subsec:MI-Entropies-with-Toeplitz}
will enable us to write R\'enyi negativities in terms of Toeplitz
determinants, and to subsequently (as was done in Subsec.~\ref{subsec:MI-Fisher-Hartwig-asymptotics})
extract their asymptotics from the Fisher-Hartwig formula.

More explicitly, we observe that the joint moments can be written
in the following integral form (again, identifying $k_{0}=k_{p}$)
\citep{fraenkel2022extensive}:
\begin{equation}
{\rm Tr}\!\left[C_{\gamma_{1}}\ldots C_{\gamma_{p}}\right]=\int_{\left[-k_{F,R},k_{F,L}\right]^{p}}\!\!\frac{d^{p}k}{\left(2\pi\right)^{p}}\prod_{j=1}^{p}\!\left[\left(1-e^{\frac{2\pi i\gamma_{j}}{n}}\right)\!\!\sum_{m\in A_{L}}\!\left\langle m|k_{j-1}\right\rangle \left\langle k_{j}|m\right\rangle +\left(1+e^{\frac{-2\pi i\gamma_{j}}{n}}\right)\!\!\sum_{m\in A_{R}}\!\left\langle m|k_{j-1}\right\rangle \left\langle k_{j}|m\right\rangle \right].\label{eq:Renyi-negativity-integral-decomposition}
\end{equation}
As in \ref{subsec:MI-Entropies-with-Toeplitz}, we assume for concreteness
that $k_{{\scriptscriptstyle F,L}}>k_{{\scriptscriptstyle F,R}}$,
and examine a simplified state where only scattering states associated
with $k_{{\scriptscriptstyle F,R}}\le k\le k_{{\scriptscriptstyle F,L}}$
are occupied (implying that the integration domain in Eq.~(\ref{eq:Renyi-negativity-integral-decomposition})
becomes $\left[k_{{\scriptscriptstyle F,R}},k_{{\scriptscriptstyle F,L}}\right]^{p}$),
and where the scattering matrix is taken to be $k$-independent. Then,
we employ the same stationary phase approximation used to derive Eq.~(\ref{eq:Approximated-correlation-moment})
(see Appendix \ref{sec:Stationary-phase-approximation} for details),
and replace the integral with a Riemann sum corresponding to the division
of $\left[k_{{\scriptscriptstyle F,R}},k_{{\scriptscriptstyle F,L}}\right]$
into $M\gg1$ equal-length discretization intervals (as in Eq.~(\ref{eq:Riemann-sum-approximated-moment})).
This allows us to write the $n$th R\'enyi negativity as 
\begin{equation}
{\cal E}_{n}\approx\sum_{\gamma=-\frac{n-1}{2}}^{\frac{n-1}{2}}\ln\widetilde{{\cal Z}}_{\gamma},\label{eq:Renyi-negativity-as-log_det-sum}
\end{equation}
where $\widetilde{{\cal Z}}_{\gamma}=\det\widetilde{K}_{\gamma}$
are determinants of $M\times M$ Toeplitz matrices, the entries of
which can be written using appropriate Toeplitz symbols as 
\begin{equation}
\left(\widetilde{K}_{\gamma}\right)_{ss'}=\int_{-\pi}^{\pi}\frac{d\theta}{2\pi}e^{-i\left(s-s'\right)\theta}\tilde{\phi}_{\gamma}\!\left(\theta\right).
\end{equation}
By approximating sums over site indices $m$ using integrals of a
continuous variable $\theta\in\left[-\pi,\pi\right]$ (as in Eq.~(\ref{eq:Toeplitz-symbol})),
we find that these symbols satisfy
\begin{equation}
\tilde{\phi}_{\gamma}\!\left(\theta\right)\approx\begin{cases}
\widetilde{T}_{\gamma}\!\left(\theta\right) & -\pi\le\theta<0,\\
{\cal T}\cdot\widetilde{T}_{\gamma}\!\left(\theta\right)+{\cal R}\cdot\widetilde{T}_{\gamma}\!\left(-\theta\right) & 0\le\theta<\pi,
\end{cases}\label{eq:Toeplitz-symbol-negativity}
\end{equation}
where
\begin{equation}
\widetilde{T}_{\gamma}\!\left(\theta\right)=\begin{cases}
e^{2\pi i\gamma/n} & \theta\in\left[-\theta_{{\scriptscriptstyle L},+},-\theta_{{\scriptscriptstyle L},-}\right],\\
-e^{-2\pi i\gamma/n} & \theta\in\left[\theta_{{\scriptscriptstyle R},-},\theta_{{\scriptscriptstyle R},+}\right],\\
1 & \text{otherwise}\,.
\end{cases}
\end{equation}
The discontinuity points $\theta_{i,\pm}$ were already defined in
Eq.~(\ref{eq:Toeplitz-symbol-discontinuities}).

Now, in analogy to Eq.~(\ref{eq:Fisher-Hartwig-asymptotics-simplified}),
the Fisher-Hartwig asymptotic formula (for large $M$) yields
\begin{equation}
\ln\widetilde{{\cal Z}}_{\gamma}\sim\frac{1}{2\pi}\left[\sum_{r}\left(\theta_{r+1}-\theta_{r}\right)\ln\!\left(\tilde{\phi}_{\gamma}\!\left(\theta_{r}^{+}\right)\right)\right]M-\frac{1}{2\pi^{2}}\sum_{r_{1}<r_{2}}\ln\!\left(\frac{\tilde{\phi}_{\gamma}\!\left(\theta_{r_{1}}^{-}\right)}{\tilde{\phi}_{\gamma}\!\left(\theta_{r_{1}}^{+}\right)}\right)\ln\!\left(\frac{\tilde{\phi}_{\gamma}\!\left(\theta_{r_{2}}^{-}\right)}{\tilde{\phi}_{\gamma}\!\left(\theta_{r_{2}}^{+}\right)}\right)\ln\!\left|M\left(e^{i\theta_{r_{2}}}-e^{i\theta_{r_{1}}}\right)\right|,\label{eq:Fisher-Hartwig-asymptotics-negativity}
\end{equation}
where in general the symbol $\tilde{\phi}_{\gamma}$ has 6 discontinuity
points $\theta_{r}$ (except for degenerate cases, where it has 4
or 5 discontinuities), but in the long-range limit $d_{i}/\ell_{i}\to\infty$
those at $\theta=-\theta_{{\scriptscriptstyle L},\pm}$ may be ignored
in the logarithmic term, except for when they ``interact'' with
each other. In Appendix \ref{sec:Negativity-derivation-appendix}
we show that the linear term in Eq.~(\ref{eq:Fisher-Hartwig-asymptotics-negativity})
indeed gives the leading-order term expected from Eq.~(\ref{eq:Renyi-negativities-asymptotics}),
assuming $k$-independent scattering probabilities. The logarithmic
term in Eq.~(\ref{eq:Fisher-Hartwig-asymptotics-negativity}) provides
a way to extract the first subleading correction, as was done for
the R\'enyi MI in Subsec.~\ref{subsec:MI-Fisher-Hartwig-asymptotics}.
Similar to the case there, this will again require separate calculations
depending on the relative positions of the discontinuity points $\left\{ \theta_{{\scriptscriptstyle L},\pm},\theta_{{\scriptscriptstyle R},\pm}\right\} $.
The execution of these calculations is, however, more cumbersome compared
to those required in the MI case; we therefore show the result only
for the symmetric case with $\ell_{{\scriptscriptstyle L}}=\ell_{{\scriptscriptstyle R}}$
and $d_{{\scriptscriptstyle L}}=d_{{\scriptscriptstyle R}}$, and
regard it as a proof of concept for the ability to extend the method
to include all other cases.

We carry out the calculation in detail in Appendix \ref{sec:Negativity-derivation-appendix}.
As in the case of the R\'enyi MI calculation in Subsec.~\ref{subsec:Derivation-of-MI},
the sum over $\gamma$ of the logarithmic contributions is expressed
there in the form of a single contour integral of a complex function,
which then allows to treat $n$ as a continuous parameter, and to
eventually perform the analytic continuation to $n=1$. Using the
notation $\ell\equiv\ell_{{\scriptscriptstyle L}}=\ell_{{\scriptscriptstyle R}}$,
it turns out that the $n$th R\'enyi negativity satisfies the asymptotic
scaling
\begin{equation}
{\cal E}_{n}\sim\left[\frac{\Delta k}{\pi}\ln\!\left({\cal T}^{n/2}+{\cal R}^{n/2}\right)\right]\ell+\left[2Q_{n/2}\!\left({\cal T}\right)+2Q_{n/2}\!\left({\cal R}\right)-\frac{n}{4}\right]\ln\ell.\label{eq:Negativity-asymptotics-symmetric-k-independent}
\end{equation}
Again, in order to generalize the result for the subleading logarithmic
term to include the possibility of a $k$-dependent scattering matrix,
we substitute the Fermi momenta $k_{{\scriptscriptstyle F,L}}$ and
$k_{{\scriptscriptstyle F,R}}$ into the scattering probabilities,
and take the average of the contributions from the two momenta. This
produces Eq\@.~(\ref{eq:Renyi-negativity-full-asymptotics-symmetric})
which was presented as part of our main results, and which is verified
numerically in Subsec.~\ref{subsec:Comparison-to-numerics}.

\section{Discussion and outlook\label{sec:Discussion}}

In this paper we continued our study, initiated in Ref.~\citep{fraenkel2022extensive},
of the mutual information and fermionic negativity of voltage-biased
free fermions in the presence of a noninteracting impurity. We refined
the exact asymptotics of these correlation measures by deriving the
subleading logarithmic corrections to the volume-law scaling reported
in Ref.~\citep{fraenkel2022extensive}, which arises for two subsystems
on opposite sides of the impurity. These logarithmic corrections can
become comparable to the volume-law terms when the voltage is small
but finite. In congruence with our conceptual focus on long-range
correlations, we focused on the long-range limit where the distance
of each subsystem from the impurity is much larger than its length.
In this limit, we obtained the exact form of the R\'enyi and von
Neumann MI for any configuration of the two subsystems (i.e., for
any choice of their lengths and relative distances from the impurity;
see Eqs.~(\ref{eq:Renyi-MI-full-asymptotics}) and (\ref{eq:Asymptotics_of_von_Neumann_MI})),
while for the negativity and its R\'enyi counterpart we presented
results only for the symmetric configuration (i.e., when the subsystems
are of equal length and within an equal distance from the impurity;
see Eqs.~(\ref{eq:Renyi-negativity-full-asymptotics-symmetric})
and (\ref{eq:Negativity-full-asymptotics-symmetric})). We stress,
however, that the analytical method that we introduced can be straightforwardly
applied to examine the negativity for more general subsystem configurations,
as well as both the MI and the negativity away from the long-range
limit.

Our result for the MI shows that the subleading logarithmic correction
to the leading-order extensive scaling depends on the particular choice
of the impurity only through the scattering probabilities corresponding
to the Fermi energies of the two edge reservoirs. This fits the standard
picture that relates such logarithmic terms to the structure of the
Fermi surface \citep{PhysRevX.12.031022,PhysRevLett.96.100503,PhysRevLett.105.050502}.
Furthermore, the dependence of this logarithmic correction on simple
four-point ratios, which are composed of the lengths and relative
distances of the subsystems from the impurity, is strongly reminiscent
of the universal form of the MI scaling observed in equilibrium states
of conformal field theories \citep{Calabrese_disjoint_2009}. In this
regard, we highlight the peculiar result of Eq.~(\ref{eq:Renyi-entropy-asymptotics-symmetric}),
which shows that the entropy of two intervals that are positioned
symmetrically relative to the impurity is independent of the associated
scattering matrix, and identical to the equilibrium result for the
same quantity.

These similarities hint at possible extensions of our results to nonequilibrium
states of interacting theories, either conformal \citep{Erdmenger2017}
or those admitting a Fermi liquid description. Such extensions could,
in turn, shed light on reasonable physical interpretations for the
forms of the scaling coefficients appearing in our results. For instance,
given that (as mentioned in Subsec.~\ref{subsec:Preliminary-notations})
$Q_{n}\!\left(0\right)$ is simply the scaling dimension of a specific
primary field in a free-fermion conformal field theory at equilibrium,
one can presumably envision formulating a statement saying that the
function $Q_{n}$ (featured in Eqs.~(\ref{eq:Renyi-MI-full-asymptotics})
and (\ref{eq:Renyi-negativity-full-asymptotics-symmetric}) of our
main results) generalizes that scaling dimension to a nonequilibrium
setting. A more general understanding of this sort, if indeed achievable,
would elucidate universal information carried by these scaling coefficients,
and thus remains a highly desirable goal.

The analytical approach that we put forward in Sec.~\ref{sec:Derivation-of-results}
was inspired by the methodology introduced in Ref.~\citep{PhysRevB.95.165101},
where the entanglement entropies and the fermionic negativity were
calculated for free fermions in their ground state. There, the discretization
of single-particle momentum (as in Eq.~(\ref{eq:Riemann-sum-approximated-moment}))
stemmed organically from the assumption of anti-periodic boundary
conditions, and this directly gave rise to relations of the form of
Eqs.~(\ref{eq:Renyi-entropies-as-log_determinant-sums}) and (\ref{eq:Renyi-negativity-as-log_det-sum})
between entanglement measures and Toeplitz determinants, due to the
Slater determinant structure of the ground-state wavefunction. For
the nonequilibrium steady state that we studied here, however, such
boundary conditions could not be imposed, yet we were able to circumvent
this issue through the reformulation of continuous integrals as limits
of Riemann sums. By removing this obstacle of requiring specific boundary
conditions, our analysis has extended the analytical method of Ref.~\citep{PhysRevB.95.165101}
beyond its original scope, thus illustrating that it can potentially
be applied to a larger class of problems. We indeed expect that our
method will be of practical use for similar calculations, and in particular
when considering zero-temperature critical 1D models, where logarithmic
terms generically constitute the leading-order contribution to correlation
and entanglement measures.

The findings that we have presented here and in Ref.~\citep{fraenkel2022extensive}
already establish that a relatively simple nonequilibrium steady state
of a quantum many-body system exhibits a unique correlation structure.
Naturally, they call for further studies into more refined measures
-- such as symmetry-resolved entanglement measures \citep{PhysRevLett.120.200602,PhysRevA.98.032302,PhysRevB.100.235146,Bonsignori_2019,Fraenkel_2020,10.21468/SciPostPhys.8.3.046,Capizzi_2020,PhysRevB.101.235169,PhysRevB.102.014455,10.21468/SciPostPhys.10.3.054,Zhao2021}
or full counting statistics \citep{doi:10.1063/1.3149497,10.21468/SciPostPhys.8.3.036,bertini2023nonequilibrium}
-- applied to disconnected intervals on opposite sides of the impurity,
which may uncover an even richer picture. The analytical technique
we employed here is in principle very well-suited for such studies.
Additionally, we believe that these results should inspire similar
explorations into long-range quantum correlations in a wider class
of nonequilibrium steady states, including other types of inhomogeneities,
of bulk models, or of external biases. In this context, we derived
exact results for the MI and the negativity also for the case where
the edge reservoirs are at finite temperatures; these results will
be reported elsewhere \citep{FuturePublication}. One may similarly
wonder about the potential effects brought about by decoherence (at
the impurity or at the edges) \citep{PhysRevResearch.2.043052,10.21468/SciPostPhys.12.1.011},
by integrability breaking \citep{PhysRevB.98.235128,Bastianello_2019,PhysRevLett.125.070605},
or by the breaking of charge conservation. All of these questions
mark the starting points of appealing paths for future research.

\section*{Acknowledgments}

We are grateful for useful discussions with P.~Calabrese, V.~Eisler,
and E.~Sela. Our work was supported by the Israel Science Foundation
(ISF) and the Directorate for Defense Research and Development (DDR\&D)
grant No.~3427/21, by the ISF grant No.~1113/23, and by the US-Israel
Binational Science Foundation (BSF) grant No.~2020072. S.F.~thanks
the Azrieli Foundation Fellows program for their support.

\appendix

\section{\label{sec:Numerics-comparison-appendix}Comparison of analytical
results to numerics in an asymmetric subsystem configuration}

\begin{figure}
\begin{centering}
\includegraphics[viewport=10bp 50bp 1107bp 770bp,clip,width=1\columnwidth]{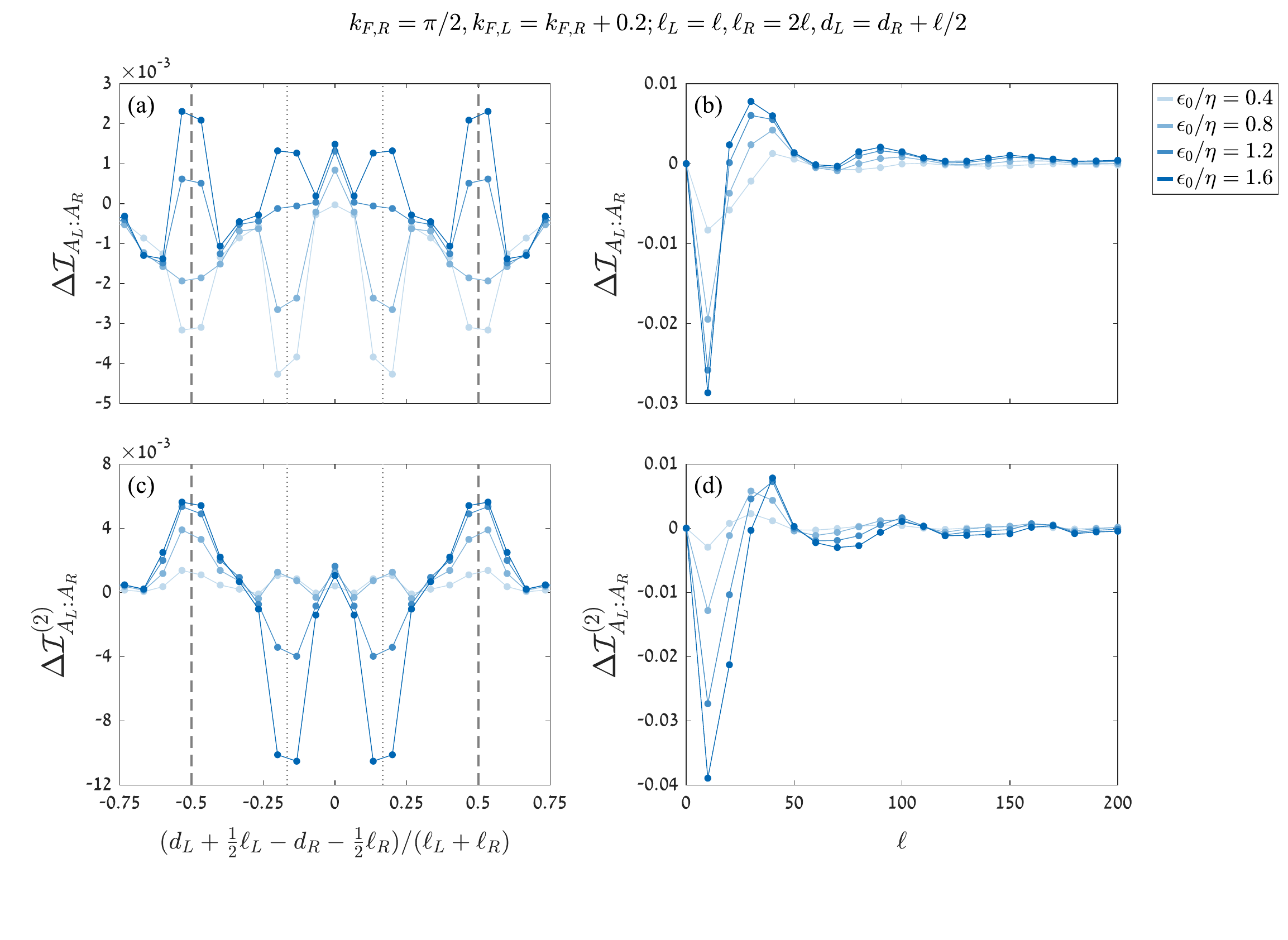}
\par\end{centering}
\caption{\label{fig:Error-of-MI-asymmetric} Deviation between numerical and
analytical results in an asymmetric subsystem configuration. Here,
for a quantity ${\cal I}$ we define the difference $\Delta{\cal I}={\cal I}^{\left({\rm num}\right)}-{\cal I}^{\left({\rm ana}\right)}$,
where ${\cal I}^{\left({\rm num}\right)}$ and ${\cal I}^{\left({\rm ana}\right)}$
are its values as obtained through numerical and analytical computations,
respectively. In all the panels, the results are computed for the
single-site impurity model (defined in Subsec.~\ref{subsec:Comparison-to-numerics})
for various values of the impurity energy $\epsilon_{0}$, with the
Fermi momenta set at $k_{{\scriptscriptstyle F,R}}=\pi/2$ and $k_{{\scriptscriptstyle F,L}}=k_{{\scriptscriptstyle F,R}}+0.2$,
and taking the long-range limit $d_{i}/\ell_{i}\to\infty$. Dots represent
the computed deviations, while thin lines serve as guides to the eye.
(a) The deviation in Fig.~\ref{fig:Scaling-of-MI-asymmetric}(a),
which shows the mutual information between subsystems $A_{{\scriptscriptstyle L}}$
and $A_{{\scriptscriptstyle R}}$ for fixed subsystem lengths ($\ell_{{\scriptscriptstyle L}}=100$
and $\ell_{{\scriptscriptstyle R}}=200$) and varying $d_{{\scriptscriptstyle L}}-d_{{\scriptscriptstyle R}}$.
The analytical results were computed using Eq.~(\ref{eq:Asymptotics_of_von_Neumann_MI}),
and the numerical results were computed as explained in Subsec.~\ref{subsec:Correlation-matrix}.
Letting $\bar{A}_{{\scriptscriptstyle L}}=\left\{ m|-\!m\in A_{{\scriptscriptstyle L}}\right\} $
denote the mirror image of $A_{{\scriptscriptstyle L}}$, dotted vertical
lines designate the range of values of $d_{{\scriptscriptstyle L}}-d_{{\scriptscriptstyle R}}$
in which $\bar{A}_{{\scriptscriptstyle L}}\subset A_{{\scriptscriptstyle R}}$,
while dashed vertical lines mark the range in which $\bar{A}_{{\scriptscriptstyle L}}\cap A_{{\scriptscriptstyle R}}\protect\neq\phi$.
(b) The deviation in the mutual information as a function of $\ell$,
for $\ell_{{\scriptscriptstyle L}}=\ell$, $\ell_{{\scriptscriptstyle R}}=2\ell$
and $d_{{\scriptscriptstyle L}}-d_{{\scriptscriptstyle R}}=\ell/2$;
under such a choice, $\bar{A}_{{\scriptscriptstyle L}}$ is centered
around the middle of $A_{{\scriptscriptstyle R}}$ for all $\ell$.
Panels (c)--(d) show the corresponding analysis for the 2-R\'enyi
mutual information (with (c) showing the deviation in Fig.~\ref{fig:Scaling-of-MI-asymmetric}(b)),
where the analytical results were computed using Eq.~(\ref{eq:Renyi-MI-full-asymptotics}).}
\end{figure}

In this appendix we briefly discuss the deviation of the analytical
results for the mutual information and for its 2-R\'enyi counterpart
from the numerical results, plotted in Fig.~\ref{fig:Scaling-of-MI-asymmetric}
for an asymmetric subsystem configuration. Figs.~\ref{fig:Error-of-MI-asymmetric}(a)
and \ref{fig:Error-of-MI-asymmetric}(c), where we subtracted the
analytical results from the numerical results, correspond respectively
to the results shown in Figs.~\ref{fig:Scaling-of-MI-asymmetric}(a)
and \ref{fig:Scaling-of-MI-asymmetric}(b). In both cases, considerable
deviations can be seen to appear mostly near the plotted vertical
dotted and dashed lines. Near each of these vertical lines, one of
length differences appearing in the logarithmic terms of the asymptotic
scaling (see the terms in the denominators in either Eq.~(\ref{eq:Renyi-MI-log-term})
or Eq.~(\ref{eq:MI-LogTerm})) becomes ${\cal O}\!\left(1\right)$,
such that contributions beyond the reach of our asymptotic calculation
are indeed expected.

To illustrate that the results of the two computations converge to
each other in the asymptotic regime, in Figs.~\ref{fig:Error-of-MI-asymmetric}(b)
and \ref{fig:Error-of-MI-asymmetric}(d) we plotted the deviation
as a function of $\ell$ when setting $\ell_{{\scriptscriptstyle L}}=\ell$,
$\ell_{{\scriptscriptstyle R}}=2\ell$ and $d_{{\scriptscriptstyle L}}-d_{{\scriptscriptstyle R}}=\ell/2$,
such that the mirror image of $A_{{\scriptscriptstyle L}}$ is always
centered around the middle of $A_{{\scriptscriptstyle R}}$ as the
lengths of the two subsystems are varied. Figs.~\ref{fig:Error-of-MI-asymmetric}(b)
and \ref{fig:Error-of-MI-asymmetric}(d) demonstrate that the deviation
decays to zero as $\ell$ gets larger, albeit in an oscillatory fashion.
These oscillations made it difficult to extract the precise scaling
of this decay through a numerical fit. More generally, given that
the analytical results in Eqs.~(\ref{eq:Renyi-MI-full-asymptotics})
and (\ref{eq:Asymptotics_of_von_Neumann_MI}) depend on more than
one length scale, understanding the scaling of this decay with respect
to all of them together is quite challenging, and is left beyond the
scope of this paper.

\section{Stationary phase approximation for correlation matrix moments\label{sec:Stationary-phase-approximation}}

In this appendix we state a central mathematical observation on which
our analysis in Ref.~\citep{fraenkel2022extensive} relied. We describe
it briefly, in order to justify a technical step that is necessary
also for the analysis that we present in the current work; a reader
who is interested in further details establishing this result should
refer to Ref.~\citep{fraenkel2022extensive}.

The observation in question pertains to the large-$\ell$ asymptotics
of $2p$-dimensional integrals of the form

\begin{equation}
{\cal F}\!\left(\overrightarrow{\tau},\overrightarrow{\sigma}\right)=\ell^{p}\int_{\left[q_{1},q_{2}\right]^{p}}\frac{d^{p}k}{\left(2\pi\right)^{p}}\int_{\left[0,1\right]^{p}}d^{p}\xi\,f_{\overrightarrow{\tau}\!,\!\overrightarrow{\sigma}}\!\left(\overrightarrow{k}\right)\exp\!\left[i\ell\sum_{j=1}^{p}\left(\left(-1\right)^{\tau_{j-1}}\!k_{j-1}-\left(-1\right)^{\sigma_{j}}\!k_{j}\right)\xi_{j}\right],\label{eq:SPA-integral-form}
\end{equation}
where $\left[q_{1},q_{2}\right]\subset\left[-\pi,\pi\right]$ is some
arbitrary interval of momentum values, $\overrightarrow{\tau},\overrightarrow{\sigma}\in\left\{ 0,1\right\} ^{\otimes p}$,
$f_{\overrightarrow{\tau}\!,\!\overrightarrow{\sigma}}$ is some $p$-dimensional
function that is independent of $\ell$ and supported on $\left[q_{1},q_{2}\right]^{p}$,
and we identify $k_{0}=k_{p}$. Suppose that $\overrightarrow{\tau}=\overrightarrow{\sigma}$,
then, by defining new variables $\left\{ \zeta_{j}\right\} _{j=1}^{p}$
such that $\zeta_{1}=\xi_{1}$ and $\zeta_{j}=\xi_{j}-\xi_{j-1}$
for $2\le j\le p$, we obtain
\begin{equation}
{\cal F}\!\left(\overrightarrow{\sigma},\overrightarrow{\sigma}\right)=\ell^{p}\int_{\left[q_{1},q_{2}\right]^{p}}\frac{d^{p}k}{\left(2\pi\right)^{p}}\int d^{p}\zeta\,f_{\overrightarrow{\sigma}\!,\!\overrightarrow{\sigma}}\!\left(\overrightarrow{k}\right)\exp\!\left[i\ell\sum_{j=2}^{p}\left(\left(-1\right)^{\sigma_{j-1}}\!k_{j-1}-\left(-1\right)^{\sigma_{p}}\!k_{p}\right)\zeta_{j}\right].
\end{equation}
We can now identify inside the exponent a function of the $\left(2p-2\right)$-dimensional
variable $\left(k_{2},\zeta_{2},\ldots,k_{p},\zeta_{p}\right)$ that
is multiplied by $i\ell$ and that has, for fixed $k_{1}$, a stationary
point at $k_{j}=\left(-1\right)^{\sigma_{1}+\sigma_{j}}k_{1}$ and
$\zeta_{j}=0$ (for $2\le j\le p$). We apply the stationary phase
approximation (SPA) \citep{doi:10.1137/1.9780898719260} to the integral
over this variable, which produces a factor that scales as ${\cal O}\!\left(\ell^{-p+1}\right)$
for large $\ell$. The remaining integral, over $k_{1}$ and $\zeta_{1}$,
is independent of $\ell$, and in total we have \citep{fraenkel2022extensive}
\begin{equation}
{\cal F}\!\left(\overrightarrow{\sigma},\overrightarrow{\sigma}\right)\sim\ell\int_{q_{1}}^{q_{2}}\!\!\frac{dk}{2\pi}\,f_{\overrightarrow{\sigma}\!,\!\overrightarrow{\sigma}}\!\left(k,\left(-1\right)^{\sigma_{1}+\sigma_{2}}k,\ldots,\left(-1\right)^{\sigma_{1}+\sigma_{p}}k\right)+o\!\left(\ell\right).
\end{equation}
This result indicates that, to a leading order, ${\cal F}\!\left(\overrightarrow{\sigma},\overrightarrow{\sigma}\right)$
scales linearly with $\ell$ (unless the integral of $f_{\overrightarrow{\sigma}\!,\!\overrightarrow{\sigma}}$
turns out to vanish), and gives the precise expression for the leading-order
term. Additional terms that grow with $\ell$, albeit more slowly
than the linear term, can in principle appear, but they are not directly
captured by the SPA (the results of our current work, which address
logarithmic corrections to linear asymptotics, imply that such terms
in fact appear in our case of interest).

In contrast, if $\overrightarrow{\tau}\neq\overrightarrow{\sigma}$,
then the same change of variables used before shows that the integral
multiplying $\ell^{p}$ in Eq.~(\ref{eq:SPA-integral-form}) decays
at least as fast as $\ell^{-p}$, such that ${\cal F}\!\left(\overrightarrow{\tau},\overrightarrow{\sigma}\right)$
is at most constant in $\ell$ (see Ref.~\citep{fraenkel2022extensive}
for a more detailed discussion of this point). We can thus summarize
the result of the SPA analysis of the integral defined in Eq.~(\ref{eq:SPA-integral-form})
as
\begin{equation}
{\cal F}\!\left(\overrightarrow{\tau},\overrightarrow{\sigma}\right)\sim\delta_{\overrightarrow{\tau},\overrightarrow{\sigma}}\left[\ell\int_{q_{1}}^{q_{2}}\!\!\frac{dk}{2\pi}\,f_{\overrightarrow{\sigma}\!,\!\overrightarrow{\sigma}}\!\left(k,\left(-1\right)^{\sigma_{1}+\sigma_{2}}k,\ldots,\left(-1\right)^{\sigma_{1}+\sigma_{p}}k\right)+o\!\left(\ell\right)\right]+{\cal O}\!\left(\ell^{0}\right).\label{eq:SPA-intgeral-asymptotics}
\end{equation}

An important point regarding the asymptotics of ${\cal F}\!\left(\overrightarrow{\sigma},\overrightarrow{\sigma}\right)$
is that the appearance of terms that grow with $\ell$ requires the
integration domain $\left[q_{1},q_{2}\right]^{p}$ to contain a subdomain
where $\left(-1\right)^{\sigma_{j-1}}k_{j-1}=\left(-1\right)^{\sigma_{j}}k_{j}$
for all $j$, since this is the location of the stationary point giving
rise to the leading-order contribution. Indeed, if for some $j$ we
have $\left(-1\right)^{\sigma_{j-1}}k_{j-1}\neq\left(-1\right)^{\sigma_{j}}k_{j}$
throughout the entire integration domain, then the variable $\xi_{j}$
in Eq.~(\ref{eq:SPA-integral-form}) can be simply integrated out,
and thus produce a multiplicative factor proportional to $\ell^{-1}$;
the SPA analysis of the remaining integral straightforwardly leads
to the conclusion that the overall scaling of ${\cal F}\!\left(\overrightarrow{\sigma},\overrightarrow{\sigma}\right)$
is at most constant in $\ell$. This happens, for example, if $q_{1}>0$
and $\sigma_{j-1}\neq\sigma_{j}$ for some $j$. Such a case in fact
arises in the computation presented in Appendix \ref{sec:Mutual-information-derivation-appendix},
and therefore this insight turns out to be crucial for an approximation
that we perform there.

Integrals of the form of Eq.~(\ref{eq:SPA-integral-form}) are relevant
to our problem due to the following reason. In the current work, all
entropies and negativities are computed from moments of correlation
matrices (see Subsec.~\ref{subsec:Correlation-matrix}). Using Eq.~(\ref{eq:Correlation_function_integral}),
we may write any moment of a restricted correlation matrix $C_{X}$
as
\begin{equation}
{\rm Tr}\!\left[\left(C_{X}\right)^{p}\right]=\int_{\left[-k_{F,R},k_{F,L}\right]^{p}}\frac{d^{p}k}{\left(2\pi\right)^{p}}\prod_{j=1}^{p}\left[\sum_{m\in X}\left\langle m|k_{j-1}\right\rangle \left\langle k_{j}|m\right\rangle \right],\label{eq:Correlation-matrix-moment-integral}
\end{equation}
whereas a generalization to joint moments, required for the R\'enyi
negativity calculations, leads to a similar integral which is given
in Eq.~(\ref{eq:Renyi-negativity-integral-decomposition}).

Consider now the case where $X=A_{i}$ (i.e, $X$ is one of the two
intervals of interest, and hence a connected subsystem). Due to the
form of the eigenstate wavefunctions (see Eqs.~(\ref{eq:Left-scattering-states})--(\ref{eq:Right-scattering-states}))
and the appearance of sums over sites $m\in X$ in Eq\@.~(\ref{eq:Correlation-matrix-moment-integral}),
a correlation matrix moment can be written as a sum of integrals of
the form of Eq.~(\ref{eq:SPA-integral-form}) (with $\ell$ replaced
by $\ell_{i}$) if one uses the identity 
\begin{equation}
\sum_{m=r+1}^{r+\ell}\exp\!\left[im\left(k_{j-1}-k_{j}\right)\right]=\ell\,{\cal W}_{r}\!\left(\frac{k_{j-1}-k_{j}}{2}\right)\int_{0}^{1}d\xi\exp\!\left[i\ell\left(k_{j-1}-k_{j}\right)\xi\right],\label{eq:Sum-as-integral-identity}
\end{equation}
where ${\cal W}_{r}\!\left(x\right)=\frac{1}{\sin x}xe^{i\left(2r+1\right)x}$
(a function which is, importantly, independent of $\ell$, and can
thus be absorbed into $f_{\overrightarrow{\tau}\!,\!\overrightarrow{\sigma}}$).
The SPA machinery described above can then be applied to obtain the
asymptotics of the correlation matrix moment.

If, on the other hand, $X=A$ (or if one considers the joint moment
of Eq.~(\ref{eq:Renyi-negativity-integral-decomposition}), in relation
to the negativity calculation), then a slight generalization of Eq.~(\ref{eq:SPA-integral-form})
is necessary to capture the form of the integrals produced by the
correlation matrix moment. Indeed, suppose that $\Delta\ell_{{\scriptscriptstyle L}}$,
$\Delta\ell_{{\scriptscriptstyle R}}$ and $\ell_{{\rm mirror}}$
all scale linearly with respect to the same large parameter $\ell$.
Then, if one again uses Eq.~(\ref{eq:Sum-as-integral-identity}),
one sees that in this case the moment is a sum of integrals of the
form
\begin{equation}
{\cal F}_{\overrightarrow{\alpha}\!,\!\overrightarrow{\beta}}\!\left(\overrightarrow{\tau},\overrightarrow{\sigma}\right)=\left[\prod_{j=1}^{p}\left(\alpha_{j}\ell\right)\right]\int_{\left[q_{1},q_{2}\right]^{p}}\frac{d^{p}k}{\left(2\pi\right)^{p}}\int_{\left[0,1\right]^{p}}d^{p}\xi'\,f_{\overrightarrow{\tau}\!,\!\overrightarrow{\sigma}}\!\left(\overrightarrow{k}\right)\exp\!\left[i\ell\sum_{j=1}^{p}\left(\left(-1\right)^{\tau_{j-1}}\!k_{j-1}-\left(-1\right)^{\sigma_{j}}\!k_{j}\right)\left(\alpha_{j}\xi_{j}'+\beta_{j}\right)\right],
\end{equation}
where $\alpha_{j}>0$ and $\beta_{j}\ge0$ are some fixed ratios related
to the scaling with $\ell$ of the different length scales. Changing
the variables by defining $\xi_{j}=\alpha_{j}\xi_{j}'+\beta_{j}$
leads to a similar form of the integral as in Eq.~(\ref{eq:SPA-integral-form}),
up to a modified integration domain of $\overrightarrow{\xi}$. The
SPA argument then leads to a similar bottom line: ${\cal F}_{\overrightarrow{\alpha}\!,\!\overrightarrow{\beta}}\!\left(\overrightarrow{\tau},\overrightarrow{\sigma}\right)$
can have a contribution beyond the constant-in-$\ell$ order only
if $\overrightarrow{\tau}=\overrightarrow{\sigma}$, and only if $\left[q_{1},q_{2}\right]^{p}$
contains a subdomain where $\left(-1\right)^{\sigma_{j-1}}k_{j-1}=\left(-1\right)^{\sigma_{j}}k_{j}$
for all $j$. We will now use this fact to simplify the integrals
that arise in our analysis before computing their asymptotics, by
dropping terms that certainly do not contribute to the logarithmic
order, as assured to us by this SPA argument.

Indeed, we employ the logic explained above to justify the approximation
given in Eq.~(\ref{eq:Approximated-correlation-moment}) for the
correlation matrix moment in the case of the simplified steady state
introduced in Subsec.~\ref{subsec:Simplified-steady-state}, where
only scattering states with $k_{{\scriptscriptstyle F,R}}\le k\le k_{{\scriptscriptstyle F,L}}$
are occupied and where the scattering matrix is $k$-independent.
We start from the exact integral expression for ${\rm Tr}\!\left[\left(C_{X}\right)^{p}\right]$,
given by Eq.~(\ref{eq:Correlation-matrix-moment-integral}), where
we only need to replace the integration domain with $\left[k_{{\scriptscriptstyle F,R}},k_{{\scriptscriptstyle F,L}}\right]^{p}$
to accommodate the definition of the simplified steady state. Recall
the general notation $X=X_{{\scriptscriptstyle L}}\cup X_{{\scriptscriptstyle R}}$
introduced in Subsec.~\ref{subsec:MI-Entropies-with-Toeplitz}, where
$X_{{\scriptscriptstyle L}}$ is the part of the subsystem that is
located to the left of the impurity, while $X_{{\scriptscriptstyle R}}$
is the part located to the right of the impurity. Then, using the
form of the scattering state wavefunctions given in Eqs.~(\ref{eq:Left-scattering-states})--(\ref{eq:Right-scattering-states})
while assuming that the scattering matrix is $k$-independent, we
observe that, for any $k_{{\scriptscriptstyle F,R}}\le k_{j-1},k_{j}\le k_{{\scriptscriptstyle F,L}}$,
\begin{equation}
\left\langle m|k_{j-1}\right\rangle \left\langle k_{j}|m\right\rangle =\begin{cases}
e^{im\left(k_{j-1}-k_{j}\right)}+{\cal R}e^{im\left(k_{j}-k_{j-1}\right)}+r_{{\scriptscriptstyle L}}e^{-im\left(k_{j-1}+k_{j}\right)}+r_{{\scriptscriptstyle L}}^{*}e^{im\left(k_{j-1}+k_{j}\right)} & m\in X_{{\scriptscriptstyle L}},\\
{\cal T}e^{im\left(k_{j-1}-k_{j}\right)} & m\in X_{{\scriptscriptstyle R}}.
\end{cases}\label{eq:Correlation-matrix-moment-explicit-integrand}
\end{equation}
Since the integration domain contains only $k_{j}>0$, the terms in
Eq\@.~(\ref{eq:Correlation-matrix-moment-explicit-integrand}) featuring
exponents of the form $\exp\left[\pm im\left(k_{j-1}+k_{j}\right)\right]$
contribute, at most, to the ${\cal O}\!\left(\ell^{0}\right)$ terms
of correlation matrix moments. These terms may therefore be omitted,
yielding the approximation in Eq.~(\ref{eq:Approximated-correlation-moment}).

\section{Logarithmic term of the R\'enyi mutual information from Fisher-Hartwig
asymptotics\label{sec:Mutual-information-derivation-appendix}}

In this appendix we complete the details of the derivation of the
R\'enyi MI logarithmic term, given by Eq.~(\ref{eq:MI-log-term-k-independent}).
As explained in Subsec.~\ref{subsec:MI-Fisher-Hartwig-asymptotics},
this requires to compute the terms
\begin{equation}
{\cal G}_{\gamma}^{\left({\rm log}\right)}={\cal G}_{\gamma}^{\left({\rm log},A_{L}\right)}+{\cal G}_{\gamma}^{\left({\rm log},A_{R}\right)}-{\cal G}_{\gamma}^{\left({\rm log},A\right)},
\end{equation}
where we used the notation
\begin{equation}
{\cal G}_{\gamma}^{\left({\rm log},X\right)}=-\frac{1}{2\pi^{2}}\sum_{r_{1}<r_{2}}\ln\!\left(\frac{\phi_{\gamma}^{\left(X\right)}\!\left(\theta_{r_{1}}^{-}\right)}{\phi_{\gamma}^{\left(X\right)}\!\left(\theta_{r_{1}}^{+}\right)}\right)\ln\!\left(\frac{\phi_{\gamma}^{\left(X\right)}\!\left(\theta_{r_{2}}^{-}\right)}{\phi_{\gamma}^{\left(X\right)}\!\left(\theta_{r_{2}}^{+}\right)}\right)\ln\!\left|M\!\left(\theta_{r_{2}}-\theta_{r_{1}}\right)\right|.
\end{equation}
Here each $\phi_{\gamma}^{\left(X\right)}$ is the Toeplitz symbol
defined in Eq.~(\ref{eq:Toeplitz-symbol}) and $\left\{ \theta_{r}\right\} $
denote its corresponding discontinuity points, apart from those that
possibly occur at $\theta=-\theta_{{\scriptscriptstyle L},\pm}$ (see
Eq.~(\ref{eq:Toeplitz-symbol-discontinuities}) for the definition
of $\theta_{{\scriptscriptstyle L},\pm}$ and $\theta_{{\scriptscriptstyle R},\pm}$).
A summation over the index $\gamma$ is the final step necessary to
obtain Eq.~(\ref{eq:MI-log-term-k-independent}). The term ${\cal G}_{\gamma}^{\left({\rm log}\right)}$
turns out to have a different structure depending on the relative
location and size of $\bar{A}_{{\scriptscriptstyle L}}=\left\{ m|-m\in A_{{\scriptscriptstyle L}}\right\} $,
the mirror image of $A_{{\scriptscriptstyle L}}$, with respect to
$A_{{\scriptscriptstyle R}}$. We go over the three different general
cases that cover all possibilities, assuming only that $\left\{ \theta_{{\scriptscriptstyle L},\pm}\right\} \cap\left\{ \theta_{{\scriptscriptstyle R},\pm}\right\} =\phi$
(in Subsec.~\ref{subsec:MI-Fisher-Hartwig-asymptotics} we briefly
discussed the extension to cases violating this assumption).

The first case we treat is that where either $\bar{A}_{{\scriptscriptstyle L}}\subset A_{{\scriptscriptstyle R}}$
or $A_{{\scriptscriptstyle R}}\subset\bar{A}_{{\scriptscriptstyle L}}$.
Assuming, for concreteness, that the former holds, we find that 
\begin{align}
{\cal G}_{\gamma}^{\left({\rm log}\right)} & =\left\{ \frac{\ln^{2}\!\left({\cal T}+{\cal R}e^{2\pi i\gamma/n}\right)-\left[\ln\!\left({\cal T}e^{2\pi i\gamma/n}+{\cal R}\right)-\frac{2\pi i\gamma}{n}\right]^{2}}{2\pi^{2}}\right\} \ln\!\left(\ell_{{\scriptscriptstyle L}}\Delta k\right)\nonumber \\
 & +\frac{\frac{2\pi i\gamma}{n}\ln\!\left({\cal T}e^{2\pi i\gamma/n}+{\cal R}\right)-\ln^{2}\!\left({\cal T}e^{2\pi i\gamma/n}+{\cal R}\right)}{2\pi^{2}}\ln\!\left|\frac{\left(\theta_{{\scriptscriptstyle R},+}-\theta_{{\scriptscriptstyle L},+}\right)\left(\theta_{{\scriptscriptstyle R},-}-\theta_{{\scriptscriptstyle L},-}\right)}{\left(\theta_{{\scriptscriptstyle R},+}-\theta_{{\scriptscriptstyle L},-}\right)\left(\theta_{{\scriptscriptstyle L},+}-\theta_{{\scriptscriptstyle R},-}\right)}\right|.\label{eq:Gamma-log-summand-full-containment}
\end{align}
For the opposite case $A_{{\scriptscriptstyle R}}\subset\bar{A}_{{\scriptscriptstyle L}}$,
${\cal G}_{\gamma}^{\left({\rm log}\right)}$ will be the same up
to replacing ${\cal T}\leftrightarrow{\cal R}$ and $L\leftrightarrow R$.
The subsequent summation over $\gamma$ can be done by observing that
the discrete sum can be expressed as a contour integral in the complex
plane, via the residue theorem. Indeed, recall the definition of the
polynomial $p_{n}$ given in Eq.~(\ref{eq:MI-polynomial-definition}).
Then, by defining a closed contour $C$ that encompasses all the roots
of $p_{n}$, we may write
\begin{equation}
\sum_{\gamma=-\frac{n-1}{2}}^{\frac{n-1}{2}}\frac{\ln^{2}\!\left({\cal T}e^{2\pi i\gamma/n}+{\cal R}\right)}{4\pi^{2}}=\frac{1}{4\pi^{2}}\int_{C}\frac{dz}{2\pi i}\cdot\frac{p_{n}'\!\left(z\right)}{p_{n}\!\left(z\right)}\ln^{2}\!\left(1-\frac{{\cal T}}{z}\right).\label{eq:gamma-sum-as-contour-integral}
\end{equation}
We can now continuously modify $C$ and take it to the limit where
it encompasses the entire complex plane apart from the ray $[0,\infty)$,
yielding
\begin{equation}
\sum_{\gamma=-\frac{n-1}{2}}^{\frac{n-1}{2}}\frac{\ln^{2}\!\left({\cal T}e^{2\pi i\gamma/n}+{\cal R}\right)}{4\pi^{2}}=\frac{1}{2\pi^{2}}\int_{0}^{{\cal T}}dx\,\frac{p_{n}'\!\left(x\right)}{p_{n}\!\left(x\right)}\ln\!\left(\frac{{\cal T}-x}{x}\right)=Q_{n}\!\left({\cal R}\right),\label{eq:Auxiliary-sum-1}
\end{equation}
where $Q_{n}$ was defined in Eq.~(\ref{eq:Log-scaling-kernel-overlap}).
In a similar manner, one may show that
\begin{align}
\sum_{\gamma=-\frac{n-1}{2}}^{\frac{n-1}{2}}\frac{i\gamma\ln\!\left({\cal T}e^{2\pi i\gamma/n}+{\cal R}\right)}{\pi n} & =\frac{1}{2\pi^{2}}\int_{C}\frac{dz}{2\pi i}\cdot\frac{p_{n}'\!\left(z\right)}{p_{n}\!\left(z\right)}\ln\!\left(1-\frac{1}{z}\right)\ln\!\left(1-\frac{{\cal T}}{z}\right)\nonumber \\
 & =\frac{1}{12}\left(\frac{1}{n}-n\right)+Q_{n}\!\left({\cal R}\right)-Q_{n}\!\left({\cal T}\right).\label{eq:Auxiliary-sum-2}
\end{align}
By using Eqs.~(\ref{eq:Auxiliary-sum-1}) and (\ref{eq:Auxiliary-sum-2}),
as well as the fact that
\begin{equation}
\sum_{\gamma=-\frac{n-1}{2}}^{\frac{n-1}{2}}\gamma^{2}=\frac{1}{12}\left(n^{3}-n\right),\label{eq:Square-index-sum}
\end{equation}
we find that the sum of the terms appearing in Eq.~(\ref{eq:Gamma-log-summand-full-containment})
is given by
\begin{equation}
\sum_{\gamma=-\frac{n-1}{2}}^{\frac{n-1}{2}}{\cal G}_{\gamma}^{\left({\rm log}\right)}=\left(Q_{n}\!\left({\cal T}\right)+Q_{n}\!\left({\cal R}\right)-\frac{1}{12}\left(\frac{1}{n}-n\right)\right)\ln\!\left|\frac{\left(\theta_{{\scriptscriptstyle R},+}-\theta_{{\scriptscriptstyle L},-}\right)\left(\theta_{{\scriptscriptstyle L},+}-\theta_{{\scriptscriptstyle R},-}\right)}{\left(\theta_{{\scriptscriptstyle R},+}-\theta_{{\scriptscriptstyle L},+}\right)\left(\theta_{{\scriptscriptstyle R},-}-\theta_{{\scriptscriptstyle L},-}\right)}\right|,\label{eq:MI-log-term-full-containment}
\end{equation}
an expression which is invariant when replacing ${\cal T}\leftrightarrow{\cal R}$
and $L\leftrightarrow R$, and which therefore applies both when $\bar{A}_{{\scriptscriptstyle L}}\subset A_{{\scriptscriptstyle R}}$
and when $A_{{\scriptscriptstyle R}}\subset\bar{A}_{{\scriptscriptstyle L}}$.

Next, let us address the case where $\bar{A}_{{\scriptscriptstyle L}}$
and $A_{{\scriptscriptstyle R}}$ do not overlap, i.e., $\bar{A}_{{\scriptscriptstyle L}}\cap A_{{\scriptscriptstyle R}}=\phi$.
In this case we find that 
\begin{equation}
{\cal G}_{\gamma}^{\left({\rm log}\right)}=\frac{\ln\!\left({\cal T}+{\cal R}e^{2\pi i\gamma/n}\right)\ln\!\left({\cal T}e^{2\pi i\gamma/n}+{\cal R}\right)}{2\pi^{2}}\ln\!\left|\frac{\left(\theta_{{\scriptscriptstyle R},+}-\theta_{{\scriptscriptstyle L},+}\right)\left(\theta_{{\scriptscriptstyle R},-}-\theta_{{\scriptscriptstyle L},-}\right)}{\left(\theta_{{\scriptscriptstyle R},+}-\theta_{{\scriptscriptstyle L},-}\right)\left(\theta_{{\scriptscriptstyle L},+}-\theta_{{\scriptscriptstyle R},-}\right)}\right|.
\end{equation}
Through a procedure similar to the one that yielded Eq.~(\ref{eq:Auxiliary-sum-1}),
we obtain
\begin{equation}
\sum_{\gamma=-\frac{n-1}{2}}^{\frac{n-1}{2}}\frac{\ln\!\left({\cal T}e^{2\pi i\gamma/n}+{\cal R}\right)\ln\!\left({\cal T}+{\cal R}e^{2\pi i\gamma/n}\right)}{2\pi^{2}}=\widetilde{Q}_{n}\!\left({\cal T}\right),\label{eq:Auxiliary-sum-3}
\end{equation}
where $\widetilde{Q}_{n}$ was defined in Eq.~(\ref{eq:Log-scaling-kernel-no-overlap}).
We therefore find that, for $\bar{A}_{{\scriptscriptstyle L}}\cap A_{{\scriptscriptstyle R}}=\phi$,
\begin{equation}
\sum_{\gamma=-\frac{n-1}{2}}^{\frac{n-1}{2}}{\cal G}_{\gamma}^{\left({\rm log}\right)}=\widetilde{Q}_{n}\!\left({\cal T}\right)\ln\!\left|\frac{\left(\theta_{{\scriptscriptstyle R},+}-\theta_{{\scriptscriptstyle L},+}\right)\left(\theta_{{\scriptscriptstyle R},-}-\theta_{{\scriptscriptstyle L},-}\right)}{\left(\theta_{{\scriptscriptstyle R},+}-\theta_{{\scriptscriptstyle L},-}\right)\left(\theta_{{\scriptscriptstyle L},+}-\theta_{{\scriptscriptstyle R},-}\right)}\right|.\label{eq:MI-log-term-no-overlap}
\end{equation}
The remaining case is that of partial overlap, that is, when $\bar{A}_{{\scriptscriptstyle L}}\cap A_{{\scriptscriptstyle R}}\neq\phi$
but also $A_{{\scriptscriptstyle R}}\setminus\bar{A}_{{\scriptscriptstyle L}}\neq\phi$
and $\bar{A}_{{\scriptscriptstyle L}}\setminus A_{{\scriptscriptstyle R}}\neq\phi$.
A similar treatment leads to the result
\begin{align}
\sum_{\gamma=-\frac{n-1}{2}}^{\frac{n-1}{2}}{\cal G}_{\gamma}^{\left({\rm log}\right)} & =\widetilde{Q}_{n}\!\left({\cal T}\right)\ln\!\left|\frac{\left(\theta_{{\scriptscriptstyle L},+}-\theta_{{\scriptscriptstyle L},-}\right)\left(\theta_{{\scriptscriptstyle R},+}-\theta_{{\scriptscriptstyle R},-}\right)}{\left(\theta_{{\scriptscriptstyle R},+}-\theta_{{\scriptscriptstyle L},-}\right)\left(\theta_{{\scriptscriptstyle L},+}-\theta_{{\scriptscriptstyle R},-}\right)}\right|\nonumber \\
 & +\left(Q_{n}\!\left({\cal T}\right)+Q_{n}\!\left({\cal R}\right)-\frac{1}{12}\left(\frac{1}{n}-n\right)\right)\ln\!\left|\frac{\left(\theta_{{\scriptscriptstyle L},+}-\theta_{{\scriptscriptstyle L},-}\right)\left(\theta_{{\scriptscriptstyle R},+}-\theta_{{\scriptscriptstyle R},-}\right)}{\left(\theta_{{\scriptscriptstyle L},+}-\theta_{{\scriptscriptstyle R},+}\right)\left(\theta_{{\scriptscriptstyle L},-}-\theta_{{\scriptscriptstyle R},-}\right)}\right|.\label{eq:MI-log-term-partial-overlap}
\end{align}

The expressions in Eqs.~(\ref{eq:MI-log-term-full-containment}),
(\ref{eq:MI-log-term-no-overlap}) and (\ref{eq:MI-log-term-partial-overlap})
for the logarithmic term of the R\'enyi MI are independent of $M$,
and therefore they stay the same in the $M\to\infty$ limit. Crucially,
all of them depend on $n$ in a manner that allows treating it as
a continuous parameter, thus making the analytic continuation to $n=1$
immediate. A convenient final observation is that all three results
can be encapsulated in a single formula, given in Eq.~(\ref{eq:MI-log-term-k-independent}).

\section{R\'enyi negativity (in a symmetric subsystem configuration) from
Fisher-Hartwig asymptotics\label{sec:Negativity-derivation-appendix}}

Here we complete the details missing from Subsec.~\ref{subsec:Derivation-of-negativity}
that lead from Eq.~(\ref{eq:Fisher-Hartwig-asymptotics-negativity})
to the asymptotic formula for the R\'enyi negativity in the case
of the artificial simplified steady state (the definition of the simplified
state is given in Subsec.~\ref{subsec:Simplified-steady-state}).

We begin by defining the following convenient notation, for any even
integer $n$:
\begin{equation}
\tilde{p}_{n}\!\left(z\right)=z^{n/2}+\left(1-z\right)^{n/2}=\prod_{\gamma=\frac{1}{2}}^{\frac{n-1}{2}}\left(1-\frac{z}{\tilde{z}_{\gamma}}\right).\label{eq:Negativity-polynomial-definition}
\end{equation}
Here $\tilde{p}_{n}$ is a polynomial with $n/2$ different roots
if $n=0\!\!\!\mod\!4$, and $n/2-1$ different roots if $n=2\!\!\!\mod\!4$,
and these roots $\left\{ \tilde{z}_{\gamma}\right\} $ satisfy
\begin{equation}
\left(\tilde{z}_{\gamma}\right)^{-1}=\frac{e^{2\pi i\gamma/n}+e^{-2\pi i\gamma/n}}{e^{2\pi i\gamma/n}},\,\,\,\,\,\gamma=\frac{1}{2},\frac{3}{2},\ldots,\frac{n-1}{2}.
\end{equation}
Next, by substituting the Toeplitz symbol of Eq.~(\ref{eq:Toeplitz-symbol-negativity}),
we observe that the linear term in Eq.~(\ref{eq:Fisher-Hartwig-asymptotics-negativity})
amounts to
\begin{align}
\frac{1}{2\pi}\left[\sum_{r}\left(\theta_{r+1}-\theta_{r}\right)\ln\!\left(\tilde{\phi}_{\gamma}\!\left(\theta_{r}^{+}\right)\right)\right]M & =\Delta k\left\{ i\ell_{{\scriptscriptstyle L}}\frac{\gamma}{n}+\frac{\Delta\ell_{{\scriptscriptstyle L}}}{2\pi}\ln\!\left[{\cal R}e^{2\pi i\gamma/n}+{\cal T}\right]+\frac{\Delta\ell_{{\scriptscriptstyle R}}}{2\pi}\ln\!\left[{\cal R}-e^{-2\pi i\gamma/n}{\cal T}\right]\right\} \nonumber \\
 & +\frac{\ell_{{\rm mirror}}\Delta k}{2\pi}\ln\!\left[{\cal R}e^{2\pi i\gamma/n}-e^{-2\pi i\gamma/n}{\cal T}\right].
\end{align}
By summing over the index $\gamma$ (using the decompositions of the
polynomials $p_{n}$ and $\tilde{p}_{n}$ in Eqs.~(\ref{eq:MI-polynomial-definition})
and (\ref{eq:Negativity-polynomial-definition}), respectively), we
thus find that, to the leading linear order, the $n$th R\'enyi negativity
scales as
\begin{equation}
{\cal E}_{n}\sim\frac{\Delta k}{2\pi}\left\{ \left(\Delta\ell_{{\scriptscriptstyle L}}+\Delta\ell_{{\scriptscriptstyle R}}\right)\ln\!\left[{\cal T}^{n}+{\cal R}^{n}\right]+2\ell_{{\rm mirror}}\ln\!\left[{\cal T}^{n/2}+{\cal R}^{n/2}\right]\right\} .\label{eq:Renyi-negativity-linear-scaling-term-k-independent}
\end{equation}
This indeed matches the known result from Eq.~(\ref{eq:Renyi-negativities-asymptotics}),
if one substitutes $k$-independent scattering probabilities.

As for the logarithmic term of ${\cal E}_{n}$, we again focus on
the long-range limit $d_{i}/\ell_{i}\to\infty$, and, as explained
in \ref{subsec:Derivation-of-negativity}, restrict ourselves to the
case of a symmetric configuration of the subsystems with $\ell_{{\scriptscriptstyle L}}=\ell_{{\scriptscriptstyle R}}\equiv\ell$
and $d_{{\scriptscriptstyle L}}=d_{{\scriptscriptstyle R}}$. Then,
the symbol $\tilde{\phi}_{\gamma}\!\left(\theta\right)$ has 4 discontinuities,
and for $M\to\infty$ the logarithmic term of Eq.~(\ref{eq:Fisher-Hartwig-asymptotics-negativity})
converges to
\begin{equation}
-\frac{1}{2\pi^{2}}\sum_{r_{1}<r_{2}}\ln\!\left(\frac{\tilde{\phi}_{\gamma}\!\left(\theta_{r_{1}}^{-}\right)}{\tilde{\phi}_{\gamma}\!\left(\theta_{r_{1}}^{+}\right)}\right)\ln\!\left(\frac{\tilde{\phi}_{\gamma}\!\left(\theta_{r_{2}}^{-}\right)}{\tilde{\phi}_{\gamma}\!\left(\theta_{r_{2}}^{+}\right)}\right)\ln\!\left|M\left(e^{i\theta_{r_{2}}}-e^{i\theta_{r_{1}}}\right)\right|\longrightarrow\left\{ -\frac{2\gamma^{2}}{n^{2}}+\frac{\ln^{2}\!\left[{\cal R}e^{2\pi i\gamma/n}-e^{-2\pi i\gamma/n}{\cal T}\right]}{2\pi^{2}}\right\} \ln\!\left(\ell\Delta k\right).\label{eq:Gamma-log-summand-negativity}
\end{equation}
To facilitate the summation over $\gamma$, we use the polynomial
defined in Eq.~(\ref{eq:Negativity-polynomial-definition}) to write
\begin{align}
\sum_{\gamma=-\frac{n-1}{2}}^{\frac{n-1}{2}}\frac{\ln^{2}\!\left[{\cal R}e^{2\pi i\gamma/n}-e^{-2\pi i\gamma/n}{\cal T}\right]}{2\pi^{2}} & =\frac{1}{\pi^{2}}{\rm Re}\!\left[\sum_{\gamma=\frac{1}{2}}^{\frac{n-1}{2}}\ln^{2}\!\left(i\left(\frac{\tilde{z}_{\gamma}}{\tilde{z}_{\gamma}-1}\right)^{1/2}\left(1-\frac{{\cal T}}{\tilde{z}_{\gamma}}\right)\right)\right]\nonumber \\
 & =\frac{1}{\pi^{2}}{\rm Re}\!\left[\int_{C}\frac{dz}{2\pi i}\cdot\frac{\tilde{p}_{n}'\!\left(z\right)}{\tilde{p}_{n}\!\left(z\right)}\left\{ \ln^{2}\!\left(i\left(\frac{z}{z-1}\right)^{1/2}\left(1-\frac{{\cal T}}{z}\right)\right)+\frac{\pi^{2}}{4}\right\} -\frac{n\pi^{2}}{8}\right],
\end{align}
where $C$ is a closed contour that encircles the entire complex plane
except for the segment $\left[0,1\right]$. Note that in the transition
from a finite sum to an integral, we added a term that ensures that
the integrand decays fast enough as $z\to\infty$, and that the contribution
of this term is subtracted after the integration is performed. This
then leads to
\begin{align}
\sum_{\gamma=-\frac{n-1}{2}}^{\frac{n-1}{2}}\frac{\ln^{2}\!\left[{\cal R}e^{2\pi i\gamma/n}-e^{-2\pi i\gamma/n}{\cal T}\right]}{2\pi^{2}} & =2Q_{n/2}\!\left({\cal T}\right)+2Q_{n/2}\!\left({\cal R}\right)-\frac{1}{6n}-\frac{n}{12},
\end{align}
where we used the definition of $Q_{n}$ in Eq.~(\ref{eq:Log-scaling-kernel-overlap}).
Combining this with the identity in Eq.~(\ref{eq:Square-index-sum}),
we may sum the terms in Eq.~(\ref{eq:Gamma-log-summand-negativity})
to obtain the logarithmic term appearing in Eq.~(\ref{eq:Negativity-asymptotics-symmetric-k-independent}).

\bibliography{Exact_Asymptotics_NESS}

\end{document}